\documentclass[12pt]{elsarticle}
\usepackage{xcolor}

\usepackage{verbatim}

\usepackage[utf8]{inputenc}

\usepackage{graphicx}

\usepackage{amssymb}

\usepackage[modulo]{lineno}
\modulolinenumbers[2]

\usepackage[hyphens]{url}

\usepackage[bookmarks, breaklinks, hidelinks, pdfauthor={Weiming Hu}, unicode=true]{hyperref}

\usepackage{algorithm}
\usepackage[noend]{algpseudocode}

\hypersetup{linkcolor=blue,citecolor=blue,filecolor=black}

\usepackage[a4paper, total={6.5in, 9in}]{geometry}

\usepackage{outlines}

\usepackage{textcomp}
\usepackage{gensymb}

\usepackage{enumitem}

\usepackage{amssymb}

\usepackage{pifont}

\usepackage[numbers]{natbib}
\newlist{todo}{itemize}{2}
\setlist[todo]{label=$\square$}

\usepackage{caption}
\usepackage{subcaption}

\usepackage[toc, section=section, acronym]{glossaries}
\usepackage[toc, section=section]{glossaries}

\usepackage{glossary-mcols}

\setacronymstyle{long-short}

\setglossarystyle{mcolindex}

\newcommand*{\Fig}[1]{Figure~\ref{#1}}
\newcommand*{\Figs}[1]{Figures~\ref{#1}}
\newcommand*{\Table}[1]{Table~\ref{#1}}

\newcommand*{\Eq}[1]{Equation~(\ref{#1})}

\newcommand*{\Sect}[1]{Section~\ref{#1}}

\usepackage{multirow}
\usepackage[normalem]{ulem}
\useunder{\uline}{\ul}{}

\usepackage[figuresleft]{rotating}
\newacronym{ADAM}{ADAM}{Adaptive Moment Estimation}
\newacronym{AI}{AI}{Artificial Intelligence}
\newacronym{AnEn}{AnEn}{Analog Ensemble}
\newacronym{ANN}{ANN}{Artifitial Neural Network}
\newacronym{API}{API}{Application Programming Interface}
\newacronym{ASOS}{ASOS}{Automated Surface Observing System}
\newacronym{ATSDR}{ATSDR}{Agency for Toxic Substances and Diseases Registry}
\newacronym{AVS}{AVS}{Agrivoltaic System}
\newacronym{Brier}{Brier}{Brier Score}
\newacronym{CDF}{CDF}{Cumulative Distribution Function}
\newacronym{CLI}{CLI}{Command Line Interface}
\newacronym{CMIP}{CMIP}{Coupled Model Intercomparison Project}
\newacronym{CONUS}{CONUS}{Continental United States}
\newacronym{CRM}{CRM}{Cloud Resolving Models}
\newacronym{CRMSE}{CRMSE}{Centered Root Mean Square Error}
\newacronym{CRPS}{CRPS}{Continuous Rank Probability Score}
\newacronym{CV}{CV}{Computer Vision}
\newacronym{DA}{DA}{Deep Analog}
\newacronym{EA}{EA}{Evolutionary Analog}
\newacronym{ECMWF}{ECMWF}{European Center for Medium Weather Forecasting}
\newacronym{ENSO}{ENSO}{El Ni\~no Southern Oscillation}
\newacronym{FLT}{FLT}{Forecast Lead Time}
\newacronym{GA}{GA}{Genetic Algorithm}
\newacronym{GCM}{GCM}{General Circulation Model}
\newacronym{GEFS}{GEFS}{Global Ensemble Forecast System}
\newacronym{GFS}{GFS}{Global Forecast System}
\newacronym{GHG}{GHG}{Greenhouse gas}
\newacronym{GISS}{GISS}{Goddard Institute for Space Studies}
\newacronym{GSI}{GSI}{Gridpoint Statistical Interpolation}
\newacronym{HF}{HF}{Heuristic Filter}
\newacronym{HPC}{HPC}{High-Performance Computing}
\newacronym{HRRR}{HRRR}{High-Resolution Rapid Refresh}
\newacronym{IDW}{IDW}{Inverse Distance Weighted}
\newacronym{IPCC}{IPCC}{International Panel on Climate Change}
\newacronym{IS}{IS}{Independent Search}
\newacronym{ITCZ}{ITCZ}{Intertropical Convergence Zone}
\newacronym{KF}{KF}{Kalman Filter}
\newacronym{LBR}{LBR}{Land-Based Renewables}
\newacronym{LOO}{LOO}{Leave-One-Out}
\newacronym{LSTM}{LSTM}{Long Short-Term Memory}
\newacronym{MAE}{MAE}{Mean Absolute Error}
\newacronym{ME}{ME}{Mean Error}
\newacronym{ML}{ML}{Machine Learning}
\newacronym{MOS}{MOS}{Model Output Statistics}
\newacronym{MRE}{MRE}{Missing Rate Error}
\newacronym{MSE}{MSE}{Mean Square Error}
\newacronym{NAM}{NAM}{North American Mesoscale Model}
\newacronym{NCAR}{NCAR}{National Center for Atmospheric Research}
\newacronym{NCEP}{NCEP}{National Centers for Environmental Prediction}
\newacronym{NN}{NN}{Neural Network}
\newacronym{NOAA}{NOAA}{National Oceanic and Atmospheric Agency}
\newacronym{NWP}{NWP}{Numerical Weather Prediction}
\newacronym{OMT}{OMT}{Over Max Time}
\newacronym{PAN}{PAN}{Persistence Analog}
\newacronym{PAnEn}{PAnEn}{Parallel Analog Ensemble}
\newacronym{PDF}{PDF}{Probability Distribution Function}
\newacronym{PDO}{PDO}{Pacific Decadal Oscillation}
\newacronym{PEF}{PEF}{Parallel Ensemble Program}
\newacronym{PNA}{PNA}{Pacific/North American}
\newacronym{PV}{PV}{Photovoltaic}
\newacronym{PWS}{PWS}{Private Weather Station}
\newacronym{RAM}{RAM}{Random Access Memory}
\newacronym{RC}{RC}{Rank Correlation}
\newacronym{RCP}{RCP}{Representative Concentration Pathway}
\newacronym{RMSE}{RMSE}{Root Mean Square Error}
\newacronym{RPS}{RPS}{Rank Probability Score}
\newacronym{SAM}{SAM}{System Advisor Model}
\newacronym{SOM}{SOM}{Self-Organizing Map}
\newacronym{SS}{SS}{Schaake Shuffle}
\newacronym{SSE}{SSE}{Search Space Extension}
\newacronym{SSI}{SSI}{Spectral Statistical Interpolation}
\newacronym{SURFRAD}{SURFRAD}{Surface Radiation Budget}
\newacronym{SVI}{SVI}{Social Vulnerability Index}
\newacronym{SVM}{SVM}{Support Vector Machine}
\newacronym{TAU}{TAU}{Tuning and Analysis Utilities}
\newacronym{UHI}{UHI}{Urban Heat Island}
\newacronym{UMAP}{UMAP}{Uniform Manifold Approximation and Projection}
\newacronym{UML}{UML}{Unified Modeling Language}
\newacronym{USSE}{USSE}{Utility-Scale Solar Energy}
\newacronym{VGI}{VGI}{Volunteered Geographic Information}
\newacronym{WHO}{WHO}{World Health Organization}
\newacronym{WRCP}{WRCP}{World Climate Research Programme}
\newacronym{WRF}{WRF}{Weather Research and Forecasting}
\newacronym{WU}{WU}{Weather Underground}

\title{Weather Analogs with a Machine Learning Similarity Metric for \\Renewable Resource Forecasting}

\author[1]{Weiming Hu\corref{correspondingAuthor}}
\cortext[correspondingAuthor]{Corresponding author. \url{weiming@psu.edu}. 205 Walker Building, University Park, PA, 16802. \url{http://geolab.psu.edu/}}

\author[1]{Guido Cervone}
\author[2]{George Young}
\author[3]{Luca Delle Monache}

\address[1]{Department of Geography and Institute of Computational and Data Sciences, \\Pennsylvania State University}
\address[2]{Department of Meteorology and Atmospheric Science, Pennsylvania State University}
\address[3]{Center for Western Weather and Water Extremes, Scripps Institute of Oceanography, \\University of California San Diego, University of California San Diego}

\begin{document}

\begin{abstract}
The Analog Ensemble (AnEn) technique is a technique that has been shown effective on several weather problems. Unlike previous weather analogs that are sought within a large spatial domain and an extended temporal window, AnEn strictly confines space and time, and independently generates results at each grid point within a short time window. AnEn can find similar forecasts that lead to accurate and calibrated ensemble forecasts.

The central core of the AnEn technique is a similarity metric that sorts historical forecasts with respect to a new target prediction. A commonly used metric is a Euclidean distance function which computes the distance between a target and a historical set of forecasts using a weighted difference of all multivariate parameters, normalized by the standard deviation of the historical set. A significant difficulty using this metric is the definition of the weights for all the parameters. Generally, the AnEn methodology starts with a feature selection task, where a subset of parameters are selected, and then weights are identified using heuristics or an optimization process. While this method has been proven effective, it is expedient and error inducing. AnEn is, in fact, characterized by a systematic bias when applied to extreme events, and this bias can potentially be hard to compensate if only a limited set of historical forecasts is available.

This paper proposes a novel definition of weather analogs through a Machine Learning (ML) based similarity metric. The similarity metric uses neural networks that are trained and instantiated to search for weather analogs. This new metric allows incorporating all variables without requiring a prior feature selection and weight optimization. Experiments are presented on the application of this new metric to forecast wind speed and solar irradiance. Results show that the ML metric generally outperforms the original metric. The ML metric has a better capability to correct for larger errors and to take advantage of a larger search repository. Spatial predictions using a learned metric also show the ability to define effective latent features that are transferable to other locations.
\end{abstract}

\maketitle

\section{Introduction}

\gls{AnEn} \cite{delle_monache_probabilistic_2013} is a technique to generate ensemble predictions from a deterministic \gls{NWP} model and the corresponding observations. Different from its predecessors \cite{van_den_dool_new_1989, toth_long-range_1989}, \gls{AnEn} identifies weather analogs independently at each grid point \cite{sperati_gridded_2017} over a short time window, e.g., three hours, which is sometimes referred to as the \gls{IS} \gls{AnEn} \cite{Clemente2020}. \gls{AnEn} works seamlessly with high-resolution \gls{NWP} models because predictions at each model grid can be directly input into \gls{AnEn} to generate prediction ensembles. \gls{IS} allows \gls{AnEn} to be computationally efficient when deployed to supercomputers \cite{cervone_short-term_2017, hu_proceedings_2020}.

\gls{AnEn} has been shown to generate accurate and calibrated ensembles in forecasting renewable energy resources, e.g., solar irradiance and wind speed. To generate final power forecasts, these meteorological forecasts can be coupled with an additional energy simulator or can be used with the actual observed power data during the analog ensemble generation to predict power. \citeauthor{alessandrini_novel_2015} applied \gls{AnEn} to short-term \gls{PV} power forecasts at three power plants at Italy. They compared \gls{AnEn} with a quantile regression method and a persistence model and showed that \gls{AnEn} were better calibrated and had a lower error. \citeauthor{cervone_short-term_2017} investigated the performance of \gls{AnEn} plus a feed-forward neural network to further account for the physical and the engineering bias. Both \gls{MRE} and \gls{CRPS} were improved, showing better forecast accuracy and ensemble quality. \citeauthor{zhang_solar_2019} proposed a blending technique that combines weather analogs from multiple \gls{NWP} models for day-ahead \gls{PV} energy market forecast. Experiments were carried out for Southeastern Massachusetts and showed that the blending technique reduces about 60\% of the error compared to a persistence model and around 20\% compared to baseline \gls{NWP} models and a data-driven \gls{SVM} model. \citeauthor{wang_hour-ahead_2020} extended the work of \citeauthor{zhang_solar_2019} to the forecasting of \gls{PV} power ramps and frequent fluctuations in energy production within the next hour. They proposed to use meteorological forecasts from a persistence model because it has a high accuracy for very short-term forecasts and avoids the computational complexity involved with running \gls{NWP} models. Results showed that it outperformed six baseline models, reaching up to a 40\% of error reduction. The capability of \gls{AnEn} generating accurate and calibrated ensembles has also been studied and tested for wind speed forecast at the surface \cite{alessandrini_novel_2015}, in the stratosphere \cite{candido2020}, and air quality predictions \cite{dellemonache2020}. \citeauthor{vanvyve_wind_2015} and \citeauthor{shahriari_using_2020} applied \gls{AnEn} for long-term and large-scale wind energy assessment. Results showed that the uncertainty information reconstructed by \gls{AnEn} is more accurate and more reliable compared to other statistical and dynamical ensembles.

\gls{AnEn}, as it was originally defined, defines a weather similarity metric, shown in \Eq{eq:anen}. The fact that weather analogs are sought with a short temporal window at each location drastically reduces the degrees of freedom of the problem at stake. Given an adequate forecast model, \gls{AnEn} is able to find similar historical forecasts and then uses the corresponding observations to correct for forecast biases that might be present in the current model forecast. There are some key aspects of the current metric definition: (1) prior knowledge on predictors and the associated weights are needed for optimal performance. \gls{AnEn} benefits from using a few predictors to identify weather analogs and it is sensitive to predictor weights \cite{junk_predictor-weighting_2015}. Since weight optimization is a computationally expansive procedure, typically \gls{AnEn} is run using only a few predictors, although \gls{NWP} models simulate hundreds. (2) \gls{AnEn} assumed a frozen model. In an operational environment, however, \gls{NWP} models are constantly subject to parameter changes. These parameterization updates are intended for better performance of the \gls{NWP} model, but they might break the assumption of a static model and, as a result, \gls{AnEn} might not yield the expected improvement when a longer search period (e.g., spanning several years) is used.

The paper seeks to address the above issues by proposing an \gls{ML} based similarity metric, inspired by the recent progress in target detecting and face recognition in \gls{CV}. Face recognition systems are able to associate a face with an identity. The key idea is that, instead of matching two face images on a pixel basis, high-level features from face images are first extracted, such as eyes and noses. These features are then compared against all the features present in a database. If a satisfactory match can be found, the associated identification in the database is queried and used as the output of the system. Presumably, there are thousands of features one can define to describe a person's face, e.g., the skin color, the curvature of eyebrows, and the shape of the nose. However, the features used by an \gls{ML} algorithm are optimized to separate images of different identities and cluster images of the same identity. These features are usually learnable, and the approach to building effective facial features is model training. A more in-depth review on this matter is provided in \Sect{sect:model-arch}.

\gls{NN} has been used for power forecasting \cite{khodayar_robust_2015, qu_xiaoyun_short-term_2016, gensler_deep_2016, qing_hourly_2018}. In this work, we extend the original \gls{AnEn} by using a non-linear similarity metric driven by a deep embedding network, namely an \gls{LSTM} network structure. Instead of using \gls{NN} as an end-to-end modeling technique as in the contributions mentioned above for power, here we seek to refine the \gls{AnEn} forecast generation with a more powerful metric characterized by a group of trained equations rather than a single equation to account for the high level of complexity in \gls{NWP} models. Building effective weather features requires model training and therefore, model training requires known similar and dissimilar weather patterns as targets. We propose a reverse analog technique to automatically select forecast pairs that could be used during the model training.

The rest of the paper is organized as follows: \Sect{sect:method} introduces, in order, the original \gls{AnEn}, the proposed architecture with \gls{NN}, and then the reverse analogs as an effective auto-labeling technique during model training; \Sect{sect:data} describes the observational and \gls{NWP} forecast data used in this research; \Sect{sect:results} shows results of the various aspects in ensemble verification; finally, \Sect{sect:conclude} includes a summary and conclusions.

\section{Methodology}
\label{sect:method}

\subsection{Analog Ensemble}

\gls{AnEn} generates forecast ensembles from an archive of deterministic model predictions and the corresponding observations of interest. \gls{AnEn} first identifies the $M$ most similar historical forecasts to the current target forecast and then, the observations corresponding to the selected historical forecasts consist of the ensemble members. This process is repeated for each forecast cycle time (e.g., when the forecast was initiated), each forecast lead time, and each grid location independently. The parameter $M$, also called the number of analog ensemble members, is usually an integer number larger than five. The choice is often practical and depends on the size of the historical archive of predictions and observations. 

\gls{AnEn} can generate accurate and representative forecast ensembles from a deterministic forecast model. It assumes that similar weather forecasts have similar error patterns and these errors can be corrected when using observations associated with similar weather forecasts. The end forecast product, being an ensemble of historical observations, is able to compensate for the random bias, possibly associated with the observations and the underlying forecasts.

The key component of \gls{AnEn} is the definition of weather similarity. \citeauthor{delle_monache_probabilistic_2013} proposed the following equation as a measure for dissimilarity:

\begin{equation}
    \|F_{t}, A_{t^{\prime}}\|=\sum_{i=1}^{N_{v}} \frac{w_{i}}{\sigma_{f_{i}}} \sqrt{\sum_{j=-\tilde{t}}^{\tilde{t}}(F_{i, t+j}-A_{i, t^{\prime}+j})^{2}},
    \label{eq:anen}
\end{equation}

where $F_{t}$ is the multivariate target forecast at the time $t$; $A_{t^{\prime}}$ is a historical multivariate analog forecast at a historical time point $t^{\prime}$; $N_v$ is the number of variables from forecasts; $w_{i}$ is the weight parameter for the forecast variable $i$ as its importance; $\sigma_{f_{i}}$ is the standard deviation of the respective variable during the historical time period; $\tilde{t}$ indicates a short time window over which the metric is computed and it equals half the number of the additional time points to consider; finally, $F_{i, t+j}$ and $A_{i, t^{\prime}+j}$ are are the values of the respective target forecast and the past analog forecast in the time window for the variable $i$.

This metric has been proven to work in various circumstances \cite{junk_predictor-weighting_2015, frediani_object-based_2017, delle_monache_probabilistic_2013, delle_monache_air_2018, alessandrini_probabilistic_2018, hu_dynamically_2019}. Its efficient implementation \cite{hu_sea_2020} and application to large scale simulation \cite{cervone_short-term_2017} have also been studied with success. \gls{AnEn} thrives in cases where historical archives of observations and forecasts are sufficient and the uncertainty information associated with a deterministic prediction is desired. It provides an accurate and computationally efficient solution to the statistical reconstruction of ensemble members.

The current definition of weather similarity, however, poses a dilemma in its optimization process. For example, most of the past literature has been involving only a handful of predictors to calculate the dissimilarity metric while, in reality, \gls{NWP} models usually provide simulations for hundreds of weather variables that characterize a wide range of the vertical profile at a certain location. Although using fewer parameters can be more permissive to finding weather analogs with lower dissimilarity metric but these analogs might not actually be the ``good analogs'' \cite{van_den_dool_new_1989, toth_long-range_1989} in a physical sense. To another extreme, if all variables have been fed into the dissimilarity calculation in favor of finding better weather analogs in the physical sense, weight optimization then becomes a significant factor that impedes its successful application. The \gls{AnEn} performance depends heavily on the parameter weights. Currently, sensitivity studies on identifying the best weight combination usually use an extensive search algorithm or a random sampling technique. The computation of the extensive search algorithm scales exponentially with the number of variables while the random sampling technique does not guarantee to give you the optimal solution. Neither these approaches are suitable for ingesting a large number of predictor variables. As a result, the process of finding weather analogs is usually confined within a few parameters, and it is subject to individual researchers to provide a good combination of weights to the best of their knowledge.

Another potential limitation is that, the current metric uses a linear combination of weather variables to formulate the similarity metric. It is, however, likely that the response variable has a non-linear relationship with the predictor variables. Part of the non-linearity can be approximated when using the actual observations as ensemble members. But the linear assumption for defining weather analogs can still cause problems in cases of rare events. \gls{AnEn} forecasts have been found to have a systematic bias when predicting extreme events, e.g., extreme wind speed \cite{alessandrini_improving_2019} and heatwaves \cite{sidel2020heat, calovi2018gfs}. On the other hand, if the predictor variables per se are associated with a significant error, the weather analogs generated using these variables will also be compromised and they only contribute to a wrong type of ensembles. A potential solution is to consider a larger pool of predictor variables and formulate an overall non-linear similarity measure.

In the following sections, an \gls{AI} framework for identifying weather analogs combining a triplet structure and an \gls{LSTM} network is proposed. Details of the framework and the training techniques are then introduced.

\subsection{Machine Learning Model Architecture}
\label{sect:model-arch}

Computer scientists have long sought algorithms for efficient feature transformation and learning. Throughout the years, \gls{NN} have become increasingly popular and reliable for achieving impressive results on tasks like image classification and face recognition. Siamese and triplet \gls{NN}s are specifically designed to solve problems of identifying similar images and produce representative embeddings. \citeauthor{baldi_neural_1993} and \citeauthor{bromley_signature_1994} are probably the first groups to formulate the early ideas of the Siamese networks. The Siamese network is a type of \gls{NN} for tackling image identification. Instead of training a network that generates a binary output, e.g., whether two images are similar or not, the Siamese network focuses the model training on generating effective embedding vectors. These embedding vectors, also referred to as the feature vectors, characterize the original image with fewer digits. For example, an RGB image with a dimension of 128x128x3 pixels can be compressed to an embedding vector containing only 16 values. The original image representation in pixels usually has redundant information, e.g., nearby pixels having similar colors. Thus, the embedding vector seeks a efficient and compact representation of the original image.  This process is referred to as image compression or dimensionality reduction. The Siamese network is particularly effective at training embedding networks. Model weights are tuned so that similar images are placed closer to each other in a transformed space and dissimilar images are placed further away. \citeauthor{baldi_neural_1993} proposed to use the Siamese network and this type of training technique to extract features from fingerprints. \citeauthor{bromley_signature_1994} developed a similar architecture independently to solve signature verification problems and coined the term Siamese network. More recent work has focused on the effectiveness of such a learned embedding. \citeauthor{chopra_learning_2005} proposed that Siamese networks should be trained in a competitive style to encourage smaller distances between similar targets and larger distances between dissimilar targets. Human-level performance in face recognition was achieved using such an architecture \cite{taigman_deepface_2014}.

At this point, the Siamese network is able to handle two input images at a time, e.g., predicting whether two face images are of the same individual. It is, however, limited in cases where a new face image is provided and the goal is to predict whether it is of individual A or B. As a result, Siamese networks were soon extended to triplet networks. Three, rather than two, images are used during one iteration of training. The three input images are referred to as the anchor, the positive, and the negative images. The convention is that the anchor image is more similar to the positive image than to the negative image. They were designed to encourage more general and robust high-level embeddings than their predecessor because, during training, the network receives guidance not only on how two images are similar, but also how two images are different. Triplet networks have outperformed Siamese networks in generating embeddings with less contextual information \cite{hoffer_deep_2015}. Later research shown interest in both approaches, using the Siamese \cite{koch_siamese_2015, bertinetto_fully-convolutional_2016, zheng_discriminatively_2017, liu_siamese_2019} and the triplet \cite{schroff_facenet_2015, dong_triplet_2018} networks for tasks like classification and tracking.

The key idea of a triplet network is the scheme that shares model parameters during model training. During forward propagation, three images are processed and compressed but this is not accomplished with three different models. Rather, the embeddings for input images are generated using the same embedding network. This is usually referred to, in the literature, as ``copies'' of the network but, in practice, these embedding networks have the same weight during training. This scheme contributes to two important properties \cite{koch_siamese_2015}: prediction consistency and network symmetry. Prediction consistency ensures that similar inputs are mapped to a hyperspace with close vicinity and dissimilar inputs are mapped with large distances; network symmetry ensures that the order within input pairs (or triplets) will not affect the final prediction. This is important for building an effective transformation so that similar observations are indeed associated with similar weather forecasts. 

This paper follows the paradigm of a triplet network. An \gls{LSTM} network is used as the embedding network. \citeauthor{hochreiter_long_1997} proposed \gls{LSTM} to encode sequence information of arbitrary lengths and it has been successfully applied to time series prediction tasks \cite{chung_empirical_2014, qu_xiaoyun_short-term_2016, gensler_deep_2016, qing_hourly_2018, gao_day-ahead_2019}. \gls{LSTM} uses memory cells and gate units to allow information from previous time lags to flow easily into later predictions. Recall the term $\tilde{t}$ from \Eq{eq:anen} indicating a time window. This parameter is usually set to 1 but it can vary based on the application. In other words, when generating \gls{AnEn}, the length of time series could vary depending on the time window size. Therefore, the ability to encode an arbitrarily long time sequence is desired when generating embeddings for similarity calculation.

\begin{figure}
    \centering
    \includegraphics[width=0.8\textwidth]{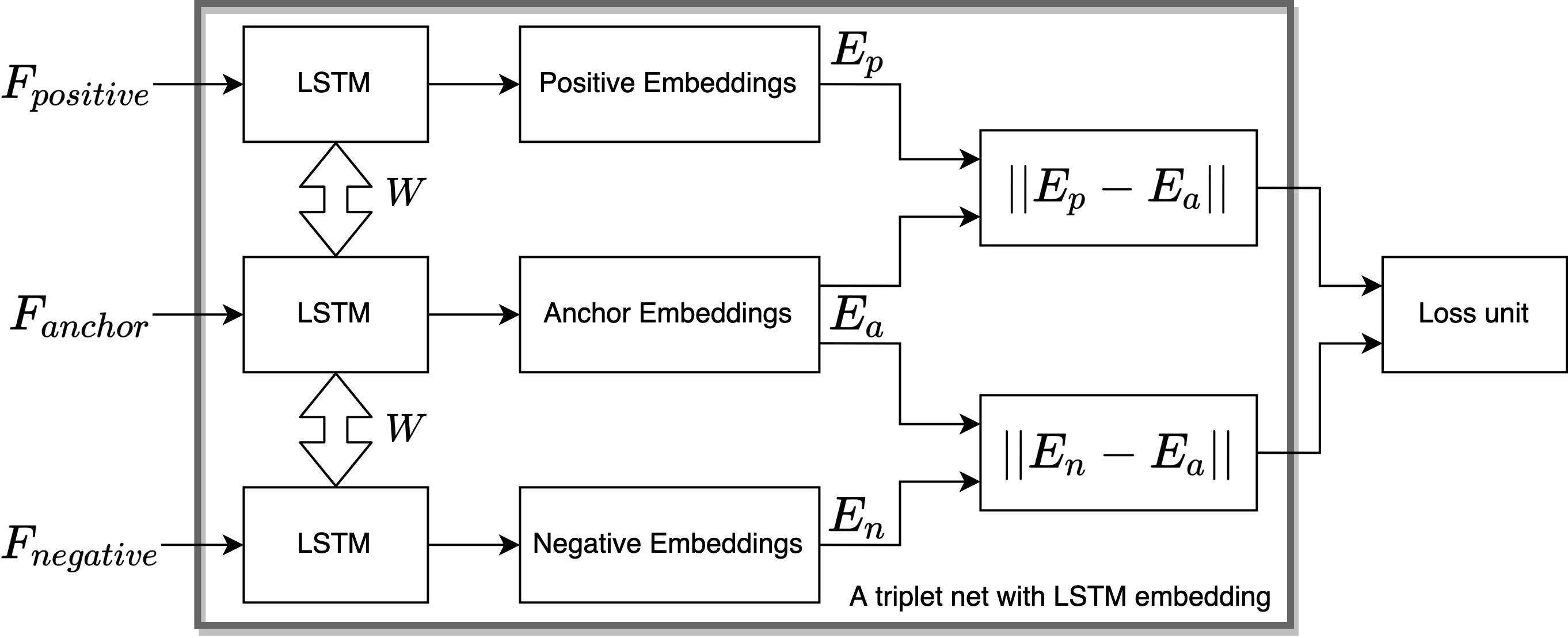}
    \caption{Triplet architecture with an \gls{LSTM} embedding network. $F$ denotes the multivariate forecast time series; $E$ denotes the embeddings; $p$ denotes the positive case; $a$ denotes the anchor case; $n$ denotes the negative case.}
    \label{fig:triplet-net}
\end{figure}

\Fig{fig:triplet-net} shows the architecture proposed in this paper, and the mathematical formulation is as follows: we seek to find a transformation function group $E_W(F)$ parameterized by $W$ and acting on a multivariate forecast time series $F$ so that, given the observations having Euclidean distances $||O_a - O_p|| < ||O_a - O_n||$, the transformation function gives $||E_W(F_a) - E_W(F_p)|| < ||E_W(F_a) - E_W(F_n)||$, where $a$, $p$, and $n$ denote the anchor, positive, and negative cases. Note that in this work, $E_W(F)$, rather than being a single function, consists of a group of functions characterized by an \gls{LSTM} network.

To explain in more details, the function group $E_W$ performs a non-linear feature transformation on the original forecast time series. The multivariate input includes a variety of variables, e.g., wind speed, temperature, and pressure. The output of this transformation is usually referred to as the embeddings, or the latent features. The latent features are usually fewer than the original input features because the latent features are intended to be more compact and abstract than the original features. More discussions about latent features are made for \Fig{fig:solar}. As mentioned before, $E_W(F)$ is a group of functions following the definition of \gls{LSTM}. It consists of the following governing equations:

\begin{enumerate}
    \item An update gate:
    \begin{equation}
        \Gamma_{u}=\sigma\left(W_{u}\left(\begin{array}{l}\mathrm{a}^{<t-1>}\\\mathrm{x}^{<t>}\end{array}\right)+b_{u}\right),
    \end{equation}
    \label{eq:update-gate}
    
    \item A forget gate:
    
    \begin{equation}
        \Gamma_{f}=\sigma\left(W_{f}\left(\begin{array}{l}\mathrm{a}^{<t-1>}\\\mathrm{x}^{<t>}\end{array}\right)+b_{f}\right),
    \end{equation}
    \label{eq:forget-gate}
    
    \item An output gate:
    \begin{equation}
        \Gamma_{o}=\sigma\left(W_{o}\left(\begin{array}{l}\mathrm{a}^{<t-1>}\\\mathrm{x}^{<t>}\end{array}\right)+b_{o}\right),
    \end{equation}
    \label{eq:output-gate}
    
    \item Input transformation:
    \begin{equation}
        \tilde{c}^{<t>}=\tanh\left(W_{c}\left(\begin{array}{l}\mathrm{a}^{<t-1>}\\\mathrm{x}^{<t>}\end{array}\right)+b_{c}\right),
    \end{equation}
    \label{eq:trans-gate}
    
    \item Cell state update:
    \begin{equation}
        c^{<t>}=\Gamma_{u} * \tilde{c}^{<t>}+\Gamma_{f} * c^{<t-1>},
    \end{equation}
    \label{eq:state-update}
    
    \item Activation output:
    \begin{equation}
        a^{<t>}=\Gamma_{o} * \tanh c^{<t>},
    \end{equation}
    \label{eq:activation}
\end{enumerate}

where $\sigma$ is a non-linear activation function for gates, usually a sigmoid function; $t$ denotes the current timestamp in the input sequence and $t-1$ denotes the previous timestamp in the input sequence; $a^{<t-1>}$ denotes the activation output from the neuron at the previous timestamp $t-1$; $x^{<t>}$ denotes the input value from the sequence at current timestamp $t$; $W$ consists of the weight parameters and $b$ denotes the linear bias term associated with a particular neuron.

\gls{LSTM} can be conceptually viewed as a combination of several \gls{NN}s. \Eq{eq:update-gate} through (4) resemble each other in that the equations are composed of first a linear transformation of the input, conditioned on parameters $W$, and second a non-linear activation function. The difference in the activation functions, $\sigma$ and $tanh$, is a practical choice in \gls{LSTM} definition. In \Eq{eq:state-update}, $\Gamma_{u} * \tilde{c}^{<t>}$ determines how much information to keep from this neuron at the current timestamp $t$; $\Gamma_{f} * c^{<t-1>}$ determines how much information to remember from the the neuron at the previous timestamp $t-1$. When $\Gamma_{u}$ is set to 0 and $\Gamma_{f}$ is set to 1, \gls{LSTM} constructs long-term memory by allowing information from previous neurons to pass without any update. The fact that \gls{LSTM} has two separate gates, $\Gamma_u$ and $\Gamma_f$, contributes to its flexibility when developing long-term and short-term memory structures. The number of neurons and hidden layers are important to \gls{LSTM}. A network with more neurons and hidden layers has more weights and is more capable of learning complex patterns. But it also takes more time and more computer memory to train and it is more prone to problems like overfitting. Choosing an appropriate configuration for the network is a cyclic procedure that involves trial and error.

The activation output, in \Eq{eq:activation}, at the last timestamp $t$ is fed into a fully connected linear layer to generate final embeddings. These embeddings are then treated as latent features to identify weather analogs using a Euclidean distance function. To identify the weights for this network, we used the \gls{ADAM} algorithm \cite{kingma2014adam} to minimize the following loss function with a learning rate of 0.005 and a dropout rate of 0.015 to prevent model overfitting:

\begin{equation}
    L = \sum_{i}^{N}[||E_W(F_a^i) - E_W(F_p^i)||-||E_W(F_a^i) - E_W(F_n^i)||+\alpha],
\end{equation}

given $||O_a^i - O_p^i|| < ||O_a^i - O_n^i||$ $\forall  i \in \{1,...,N\}$ where $N$ is the number of samples and $O^i$ is the corresponding observations for the forecast $F^i$. $E_W$ denotes the embedding \gls{LSTM} network with learned weights $W$; $||E_W(F_a^i) - E_W(F_p^i)||$ denotes the L-2 norm between the anchor and the positive forecasts for the $i$-th sample. $\alpha$ denotes a margin value and it encourages the optimization to make distances between positive pairs to be smaller than the negative pairs by this margin value.

\subsection{Machine Learning Model Training}
\label{sect:learning}

There is yet one more problem to address: how to sample triplets for model training, or in other words, how to determine $F_a$, $F_n$, and $F_p$. If triplets are sampled based on \Eq{eq:anen}, the learned model simply acts like a dimension reduction technique that approximates a sub-optimal similarity definition. No more insights are added by this learning process. Inspired by the negative sampling technique \cite{dyer2014notes, goldberg2014word2vec, xu2015semantic, wang2018incorporating}, we propose the reverse analog technique for identifying virtually optimal triplets for model training. 

\begin{figure}
    \centering
    \includegraphics[width=.65\textwidth]{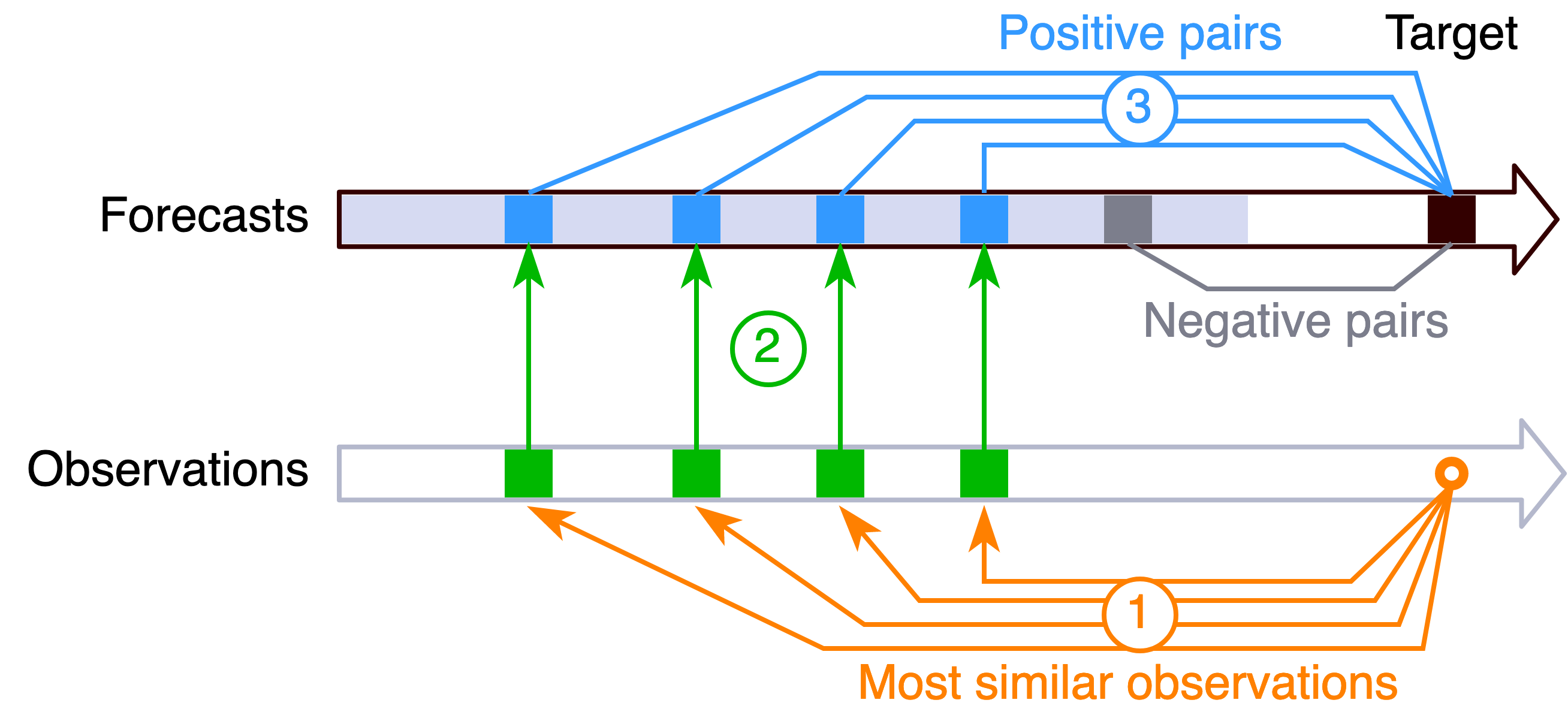}
    \caption{Schematic for identifying reverse analogs during the triplet sampling from historical forecasts and observations}
    \label{fig:reverse_analogs}
\end{figure}

\Fig{fig:reverse_analogs} shows the schematic for the reverse analog process. It echos the \gls{AnEn} schematic presented in \citeauthor{hu_dynamically_2019}'s paper but the direction of information flow has been reversed for the specific purpose of optimization. To identify which forecasts should be truly more similar than others, we refer to the corresponding observation field. In other words, similarity of forecasts is determined not from forecasts per se, but rather from the corresponding observations. If the associated observations are very similar, the corresponding forecasts are deemed analogs, although the original Euclidean distance calculated directly from forecast variables might be large. And this is indeed the additional information the triplet network aims to learn: the forecast error patterns coupled with certain types of observation values. Specifically, the reverse analog technique works as follows:

\begin{enumerate}
    \item Most similar historical observations (green rectangles) to the current observation (orange circle) are found.
    \item The associated historical forecasts (blue rectangles) are queried.
    \item The current target forecast (black rectangle) is deemed to be more similar to the matched historical forecasts (blue rectangles) than other unmatched historical forecasts (grey rectangle).
\end{enumerate}

One might argue that similar observations might not necessarily indicate actual similar weather patterns. The degree of freedom is hard to control, subject to seasonality, diurnal variability, and random variation. We offer two counterpoints in this light: (1) similar forecasts and observations are sought independently at each location and each forecast lead time which would control for spatial and diurnal variability; (2) we incorporate a fitness proportionate selection stage \cite{whitley1994genetic, hancock1994empirical} to introduce randomness and to prevent greedy search \cite{vafaie1994feature, wilt2010comparison, hu_dynamically_2019}. The goal is not to distinguish every nuance of the observational difference, but to learn an effective embedding function relating forecast errors and observations.

\section{Research Datasets}
\label{sect:data}

Experiments have been carried out at two geographic scales to show the effectiveness of using a \gls{ML} driven similarity metric.

The first geographic scale is a small scale study located at Pennsylvania State University. The ground observation station is actively maintained by the \gls{SURFRAD} project \footnote{Station coordinates can be accessed from \url{https://www.esrl.noaa.gov/gmd/grad/surfrad/sitepage.html}}. Observations of solar irradiance and surface wind speed have been collected from \gls{SURFRAD} between 2011 and 2019. \gls{SURFRAD} \cite{augustine2000surfrad, augustine2005update} project was established in 1993 to provide high-quality, continuous, and long-term measurements of the surface radiation budget. Observations from \gls{SURFRAD} have been used in various validation procedures for satellite-derived estimates and \gls{NWP} models. In our case, the verified \gls{NWP} model is the \gls{NAM} forecast system. \gls{NAM} is one of the major operational models run by \gls{NCEP} for weather predictions. It uses boundary conditions from the \gls{GFS} model and is initialized four times per day at 00, 06, 12, and 18 UTC. Each initialization produces forecasts for the next 84 hours. The first 37 lead times are hourly and then the temporal resolution is reduced to every three hours until the maximum lead time is reached. \gls{NAM} has various grid types each associated with a different spatial resolution. The outer domain with 12-km horizontal grid increments covering the \gls{CONUS} is used in this work. \gls{NAM} provides simulations for over three hundred weather variables that cover a wide range of vertical profile of the atmosphere. It simulates in total 60 vertical layers on a hybrid sigma-pressure coordinate system. It also simulates a single compound atmospheric layer for variables like downwelling shortwave solar radiation and total precipitation. \gls{NAM} forecasts \cite{environmental2015ncep} initialized at 00 UTC from 2011 to 2019 have been collected and subset to the region of Pennsylvania. The closest model grid to the \gls{SURFRAD} stations is identified. Model forecasts from this grid are verified with \gls{SURFRAD} stations.

The second geographic scale is a large scale study that covers Pennsylvania. The variables of interest are solar irradiance and wind speed at the surface and 80 meters above ground. Please note that the analysis of wind speed forecasts at 80 meters above ground is omitted in the small scale study because observations are not available from \gls{SURFRAD}, but it is included in the large scale study because it is the most relevant variable for regional wind power assessment. The \gls{NWP} model remains to be \gls{NAM} but \gls{SURFRAD} observations do not have sufficient spatial coverage of this area, so the model analysis field is used during verification. \gls{NAM} analysis initialized at 00, 06, 12, and 18 UTC are collected. Model analysis is used as a best-scenario approximation to ground truth where the ground measurements are not available. The state of Pennsylvania is chosen due to its direct closeness to the \gls{SURFRAD} station. There are in total 1225 grid points in the selected domain.

\section{Results}
\label{sect:results}

Results for wind speed and solar irradiance predictions are organized into several sections. All experiments have the same test period of 2019. The training period for the \gls{ML} model and the search period of \gls{AnEn} are both from 2011 to 2018 to ensure a fair comparison, except for \Sect{sect:model-updates} where the length of search is of particular interest. All experiments have 11 members in forecast ensembles.

There are two configurations of \gls{AnEn} that are treated as baselines. The first configuration is the optimal \gls{AnEn}. This configuration is deemed optimal in a sense that the predictors used are selected based on literature \cite{cervone_short-term_2017, delle_monache_probabilistic_2013, alessandrini_improving_2019, shahriari_using_2020, sperati_gridded_2017} and the respective weights are then optimized via an extensive search that tests for all possible combinations of weights with an increment of 0.1 from 0 to 1. The selected predictors include downwelling shortwave radiation, surface wind speed and wind direction, relative humidity and temperature at 2 meters above ground.

The second configuration is the equal weighting scheme, namely the equally weighted \gls{AnEn}. This configuration uses the same predictors as the \gls{ML} model. There are in total 227 variables selected from \gls{NAM} for both the equal weight scheme and the \gls{ML} model. Several variables are excluded from the training process due to being soil attributes, not necessarily related to atmospheric conditions. Due to the sheer number of variables, an extensive search would not be possible and therefore, equal weighting is the next most practical thing to do.

We trained two separate \gls{ML} models, one for wind speed predictions and one for solar irradiance predictions, with both following the same model architecture. The \gls{LSTM} embedding network has 20 hidden features and three hidden layers. In other words, 383 predictors are transformed into a 20-dimensional space which is then used to identify analogs. These hyper-parameters are decided based on several trials with 3, 5, 20, and 50 hidden features and 1, 2, and 3 hidden layers respectively. The maximum training iteration is 200,000 and early stopping is engaged to prevent model overfitting. In the rest of the description, predictions generated from the \gls{ML} driven similarity metric will be referred to as \gls{DA} for simplicity.

\subsection{Weight Optimization}
\label{sect:weight-optim}

\begin{figure}[ht]
    \begin{minipage}{\textwidth}
        \centering
        \includegraphics[width=\textwidth]{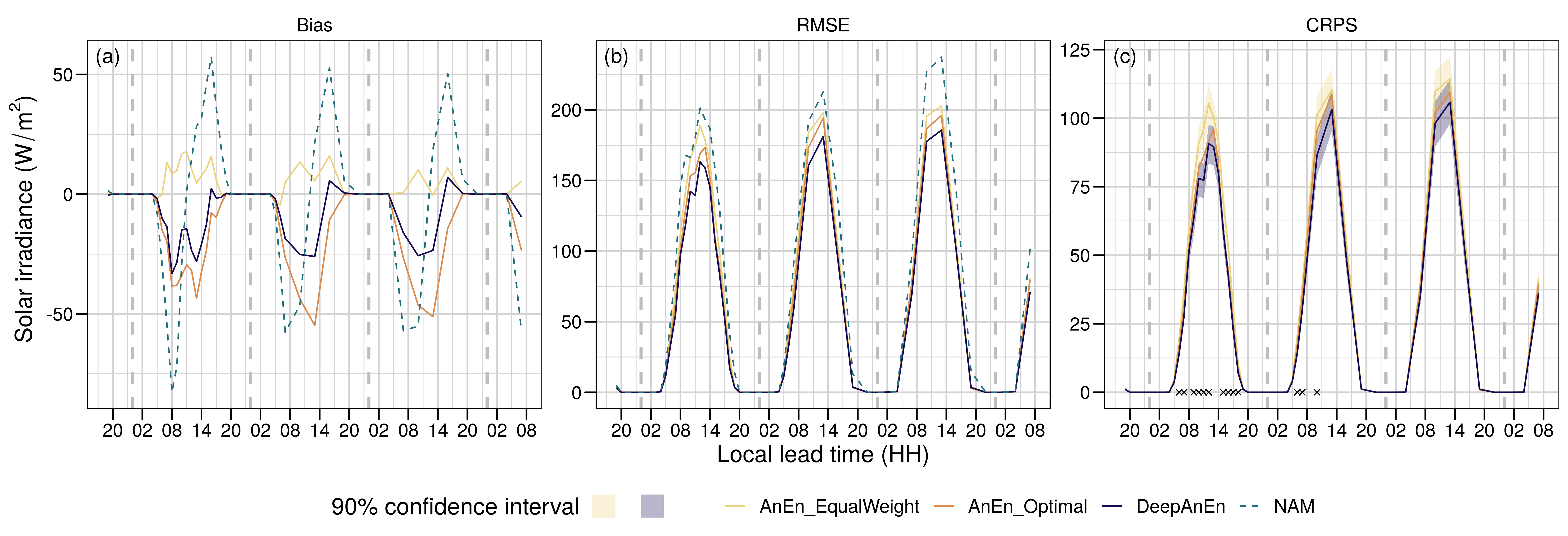}
    \end{minipage}
    \begin{minipage}{\textwidth}
        \centering
        \includegraphics[width=\textwidth]{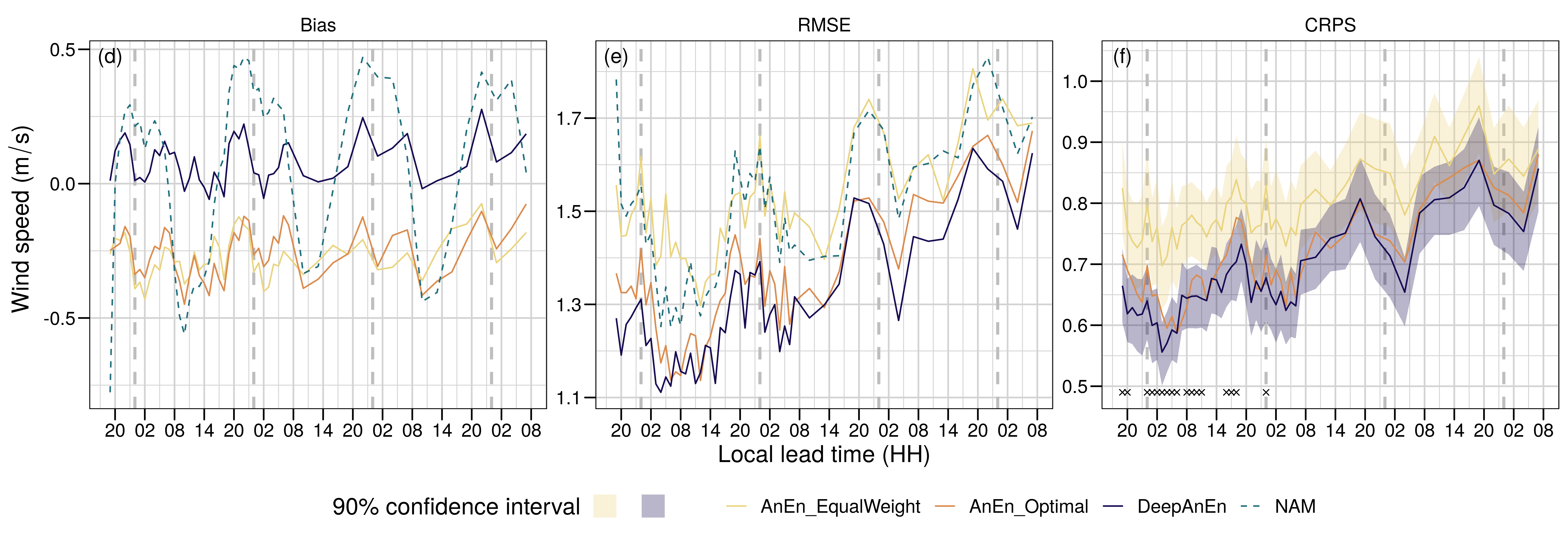}
    \end{minipage}
    \caption{Verification comparisons for (a, b, c) solar irradiance and (d, e, f) wind speed forecasts. Black crosses on the bottom of (c, f) indicate the lead time hours that \gls{DA} is significantly better than \gls{AnEn} with equal weighting at a 90\% confidence level. Vertical dashed lines indicate individual forecast days.}
    \label{fig:crps}
\end{figure}

\Figs{fig:crps} (a, b, c) show the bias, \gls{RMSE}, and \gls{CRPS} for solar irradiance forecasts. Overall, based on comparisons of \gls{RMSE} and \gls{CRPS}, \gls{DA} outperforms all other methods. Specifically, \gls{DA} significantly outperforms the equally weighted \gls{AnEn} for most of the lead times in the first 24 hours. This time period, namely the day-ahead horizon, is of particular interest in terms of the energy market bidding and it is important for energy system planning and scheduling that tries to match power demand and supply. From the bias panel, \gls{NAM} demonstrates a negative bias during mornings and a positive bias during afternoons. This suggests that \gls{NAM} has problems simulating the exact diurnal trend. All configurations of \gls{AnEn} have been able to correct this behavior but they produce slightly different results. The equally weighted \gls{AnEn} shows a positive bias while the optimal \gls{AnEn} shows a negative bias possibly caused by giving a dominant weight on downwelling shortwave radiation and then taking the average of the ensemble members. \gls{DA} achieves better bias than the optimal \gls{AnEn} for almost all lead times. This indicates that by using the transformed features from many more predictors, \gls{DA} is able to find better analogs that are missed by the optimal \gls{AnEn}. Although all configurations of \gls{AnEn} still produce bias, the bias magnitudes are very small compared to other error statistics, like \gls{RMSE} and \gls{CRPS}.

Verification on wind speed predictions demonstrates similar results, shown in \Figs{fig:crps} (d, e, f). \gls{DA} is shown to outperform both the optimal \gls{AnEn} and the \gls{AnEn} with equal weighting for most of the forecast lead times on average for \gls{RMSE} and \gls{CRPS}. \gls{DA} is significantly better, in \gls{CRPS}, than the equally weighted \gls{AnEn} for most of the lead times in the first 24 hours. These results are consistent with the verification on solar irradiance predictions. It should be noted that the \gls{RMSE} of the equally weighted \gls{AnEn} has degraded to the level of \gls{NAM} for most of the lead times when predicting wind speed. This suggests that simply increasing the number of predictor variables does not guarantee an improved accuracy. Bias-wise, \gls{DA} clearly achieves better bias than all other methods. \gls{NAM} has a similar problem, as in \Fig{fig:crps} (a), in simulating the diurnal changes in wind speed resulting to a sine wave-like bias diagram. By average, the equally weighted \gls{AnEn} has a bias of -0.271 $m/s$ and the optimal \gls{AnEn} has a bias of -0.260 $m/s$. \gls{DA} is able to reduce the bias to an average level of 0.080 $m/s$.

In terms of significance level, \gls{DA} remains significantly better than the equally weighted \gls{AnEn} for most of the lead times in the first forecast day period at a 90\% confidence level. This suggests the effectiveness of \gls{ML} model training as a weight optimization process.

\begin{figure}[htp!]
    \begin{minipage}{\textwidth}
        \centering
        \includegraphics[width=\textwidth]{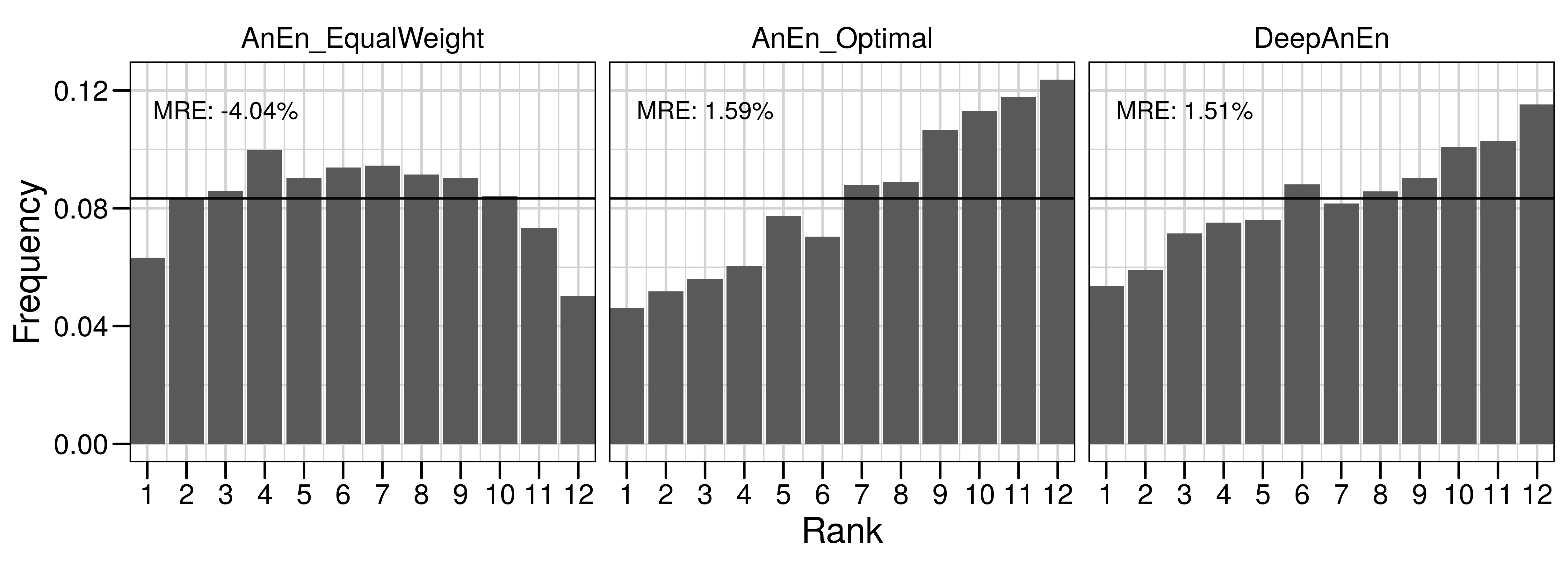}
    \end{minipage}
    \begin{minipage}{\textwidth}
        \centering
        \includegraphics[width=\textwidth]{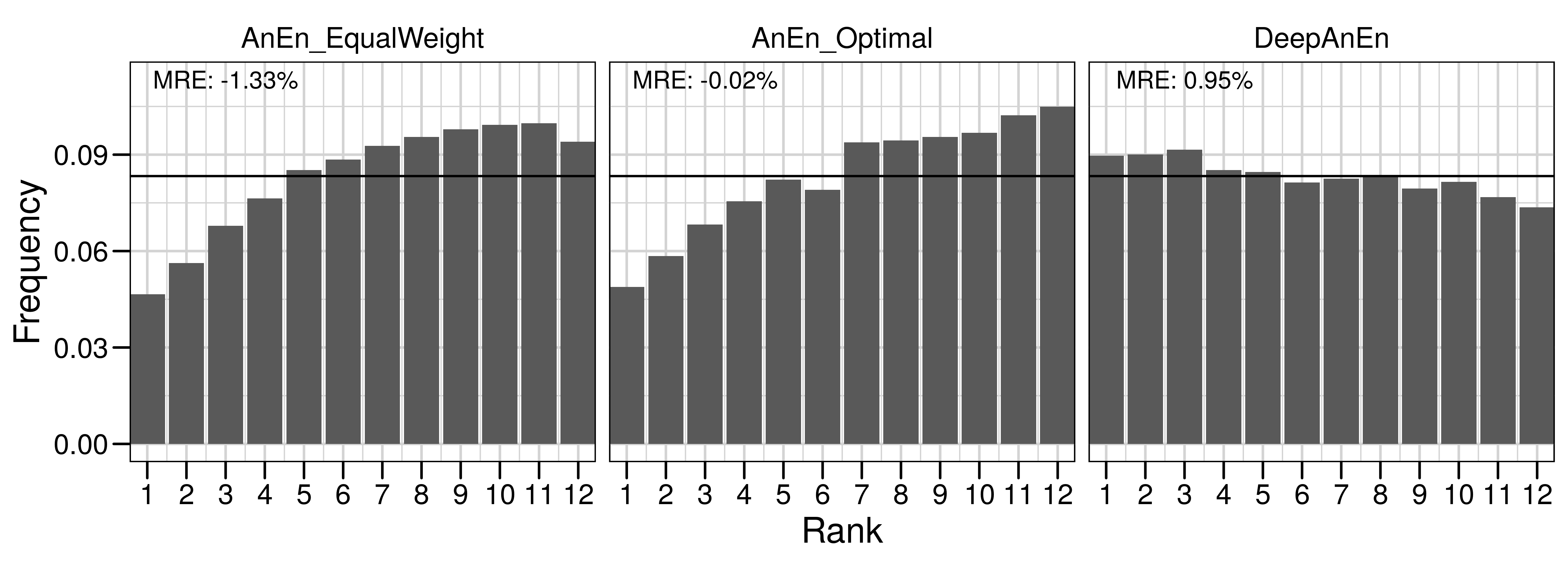}
    \end{minipage}
    \begin{minipage}{0.5\textwidth}
        \centering
        \includegraphics[width=\textwidth]{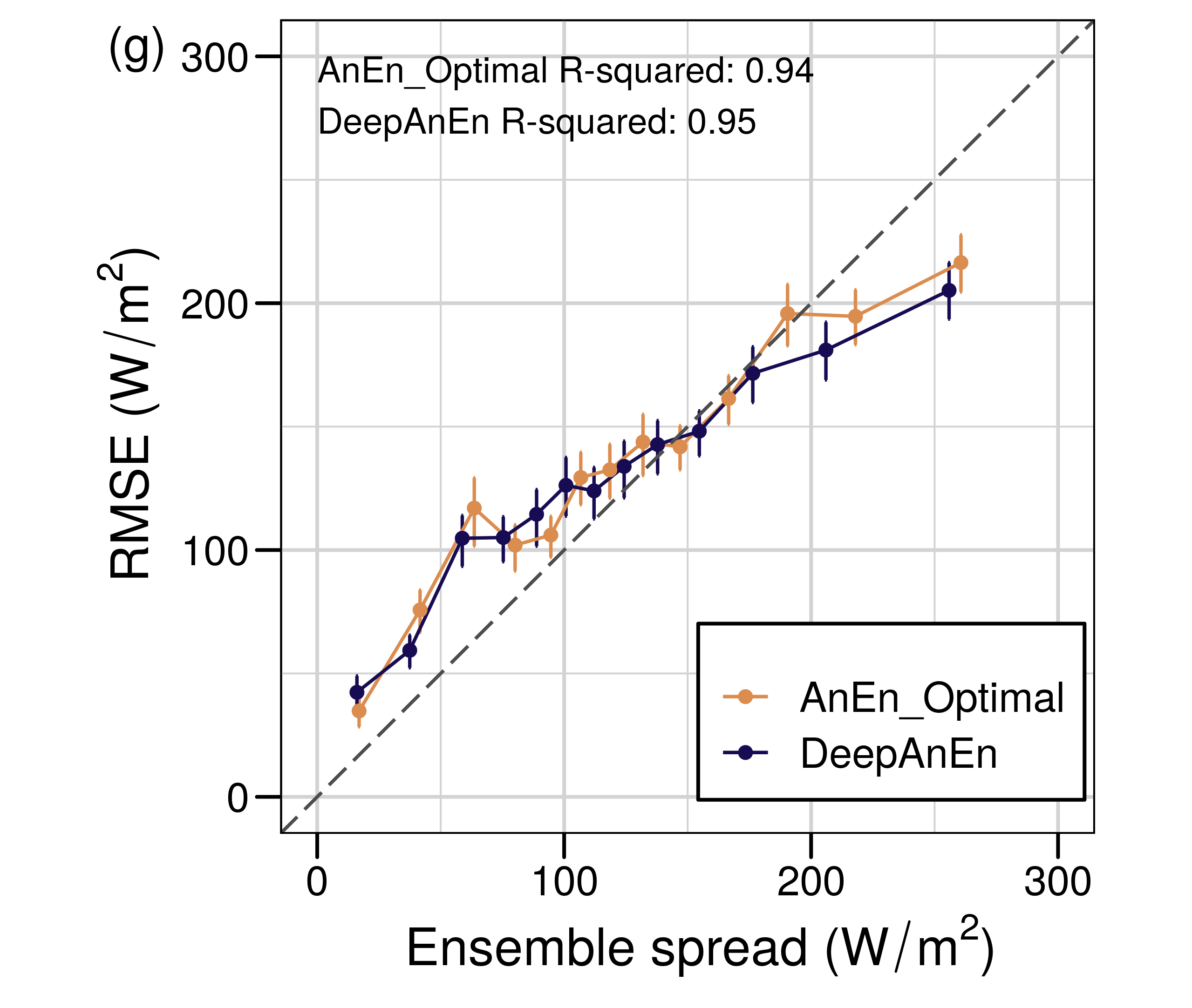}
    \end{minipage}
    \begin{minipage}{0.5\textwidth}
        \centering
        \includegraphics[width=\textwidth]{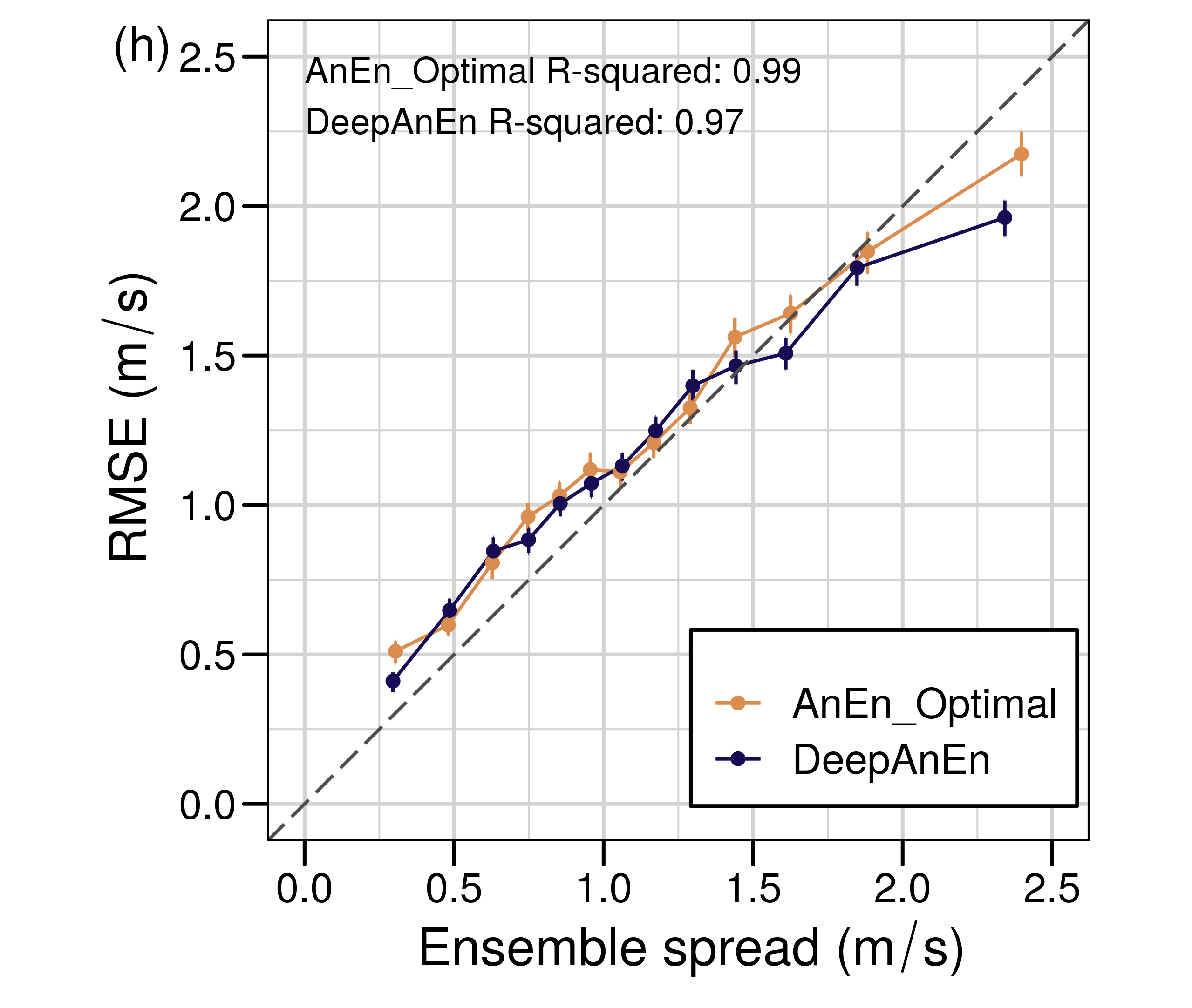}
    \end{minipage}
    \caption{Rank histograms and reliability diagrams for solar irradiance and wind speed forecasts. (a, b, c) and (d, e, f) show the rank histograms respectively for solar irradiance and wind speed forecasts with 11 ensemble members across all lead times. (g) and (h) show the binned spread-error relationship diagrams respectively for solar irradiance and wind speed forecasts. \gls{RMSE} is used as the error measure on the vertical axes in (g, h). Vertical lines on each dot indicate the 90\% confidence interval.}
    \label{fig:rh-ss}
\end{figure}

Next, the statistical consistency and the spread-skill relationship of \gls{DA} and \gls{AnEn} are evaluated with rank histograms and binned spread-error diagram, respectively, as shown in \Fig{fig:rh-ss}. \Figs{fig:rh-ss} (a, b, c) show the rank histograms for solar irradiance forecasts. The ensembles from equally weighted \gls{AnEn} is slightly over-dispersed, or under-confidence, confirmed by a negative \gls{MRE} and a convex shape of the rank histogram. When using all available variables without discrimination, weather analogs that are actually associated with the variable of interest are hard to find and as a results, the ensemble spread tends to be too large. Both \gls{DA} and the optimal \gls{AnEn} produce slightly under-dispersed ensembles as indicated by a positive \gls{MRE} while \gls{DA} is closer to zero than \gls{AnEn}. The prevailing positive slope in both rank histograms confirms the systematic low bias shown in \Fig{fig:crps} (a). \Figs{fig:rh-ss} (d, e, f) show the rank histograms for wind speed forecasts. Consistent with \Fig{fig:crps} (d), \gls{DA} successfully reduce biases in other two configurations of \gls{AnEn} by yielding an almost flat rank histogram. Its \gls{MRE} is still positive indicating ensembles being slightly under-dispersive. The ensemble characteristics of the two configurations of \gls{AnEn} are highly similar, both having a negative bias and over-dispersive ensembles.

The spread-error diagram, shown in \Figs{fig:rh-ss} (g, h), quantifies the ability of the ensemble in estimating the prediction errors. Forecasts from \gls{DA} and the optimal \gls{AnEn} have very similar ability to quantify uncertainty except for ensembles with a larger spread (right tails). \gls{DA} is able to reduce errors while hardly impacting the spread for ensembles with a spread that is larger than 180 $W/m^2$ and 1.8 $m/s$. The slightly lower R-squared measure is primarily caused by the right tails of \gls{DA} correlation deviating further from the diagonal line than the optimal \gls{AnEn}. This indicates that \gls{DA} carries out a shift to the forecasted distribution to the correct direction without changing the shape of the distribution. It is worthwhile to note that \gls{DA} achieves similar performance as the optimal \gls{AnEn} in most cases but they start to become different when looking at the hard-to-predict cases that are associated with larger errors. Some examples of such hard-to-predict cases include: (1) abrupt weather regime changes, and (2) constantly moving partial clouds. This difference will be further discussed in the following section.

\subsection{Relationship between Performances and Training Length}

Results from \Sect{sect:weight-optim} have shown that the triplet network with an \gls{LSTM} embedding is able to learn an effective transformation function from over hundreds of predictor variables down to twenty latent features. Overall, \gls{DA} outperforms the optimal \gls{AnEn}. Since the cases where differences exist between the two methods likely lies within the hard-to-predict scenarios, as shown in \Figs{fig:rh-ss} (g, h), this section focuses on comparing the ability of two methods in correcting the underlying \gls{NAM} when it is making larger errors.

\begin{figure}[htp!]
    \begin{minipage}{0.43\textwidth}
        \centering
        \includegraphics[width=\textwidth]{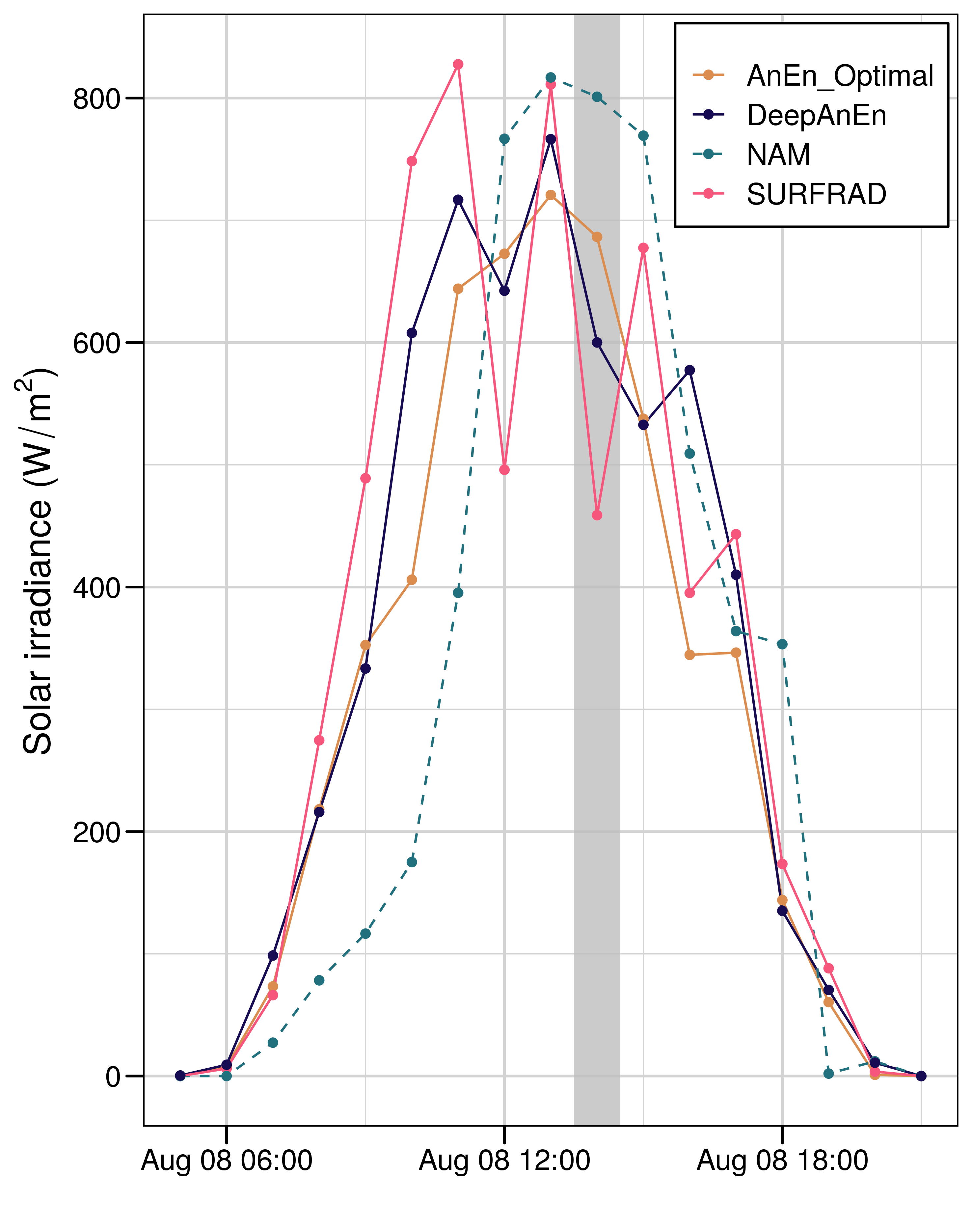}
        \phantomsubcaption
    \end{minipage}
    \begin{minipage}{0.57\textwidth}
        \centering
        \includegraphics[width=\textwidth]{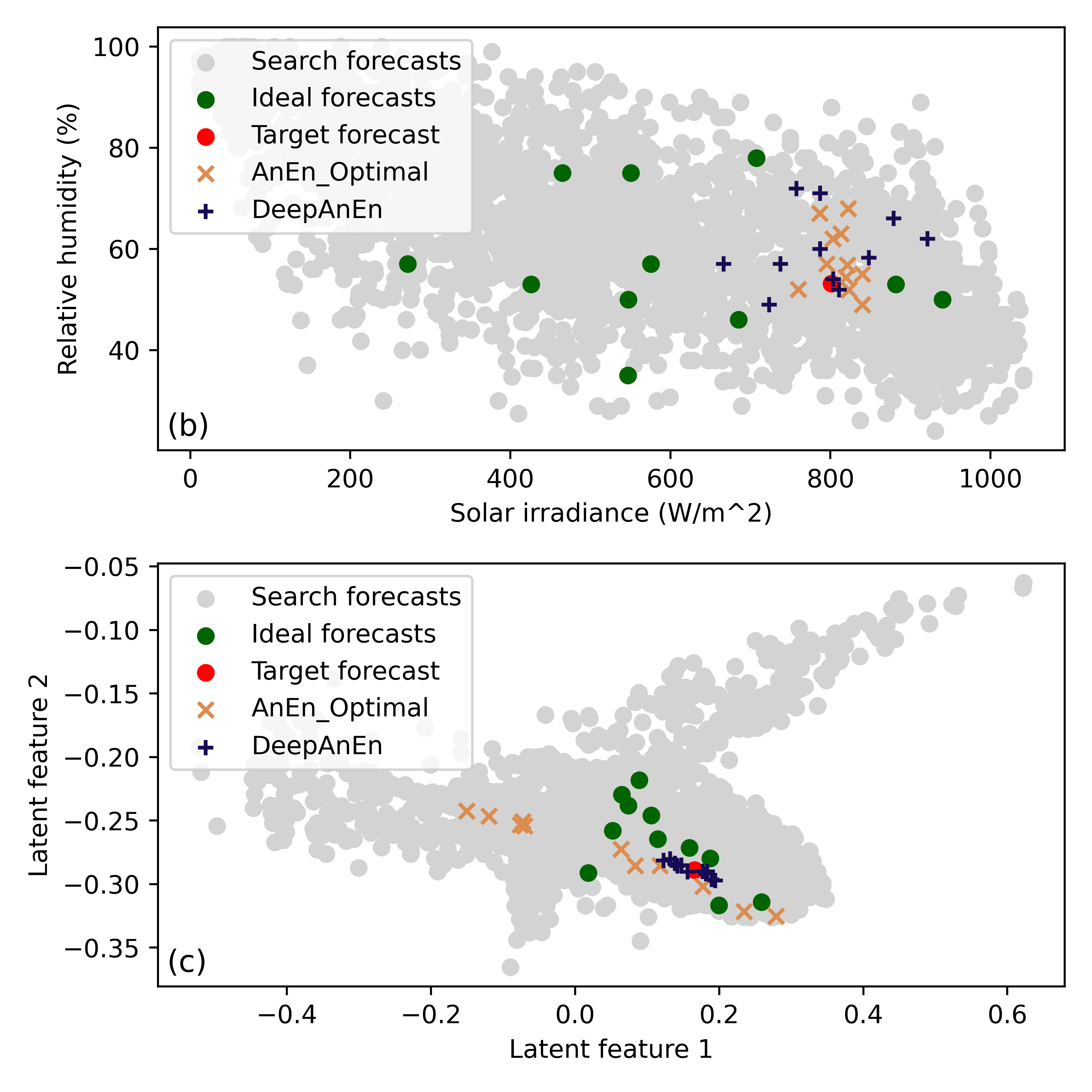}
        \phantomsubcaption
    \end{minipage}
    \caption{A case study of solar irradiance prediction on August 8th, 2019, showing the predicted time series from various methods (a). The average of ensemble members are calculated to show a single predicted time series. The shaded area is 2 PM location time. A subset of the original forecast variables and the latent features at this particular time point are shown in (b) and (c).}
    \label{fig:solar}
\end{figure}

First, let's consider a case study of solar irradiance prediction on August 8th, 2019. \Fig{fig:solar} (a) shows the observed and predicted time series from \gls{NAM}, and the ensemble mean of \gls{DA} and the optimal \gls{AnEn}. \gls{SURFRAD} time series shows a bumpy temporal trend with a large amount of variation from the late morning to the early afternoon. The average irradiance during noon, however, is around 600 $W/m^2$ which amounts to a typical semi-cloudy day. The large variation is likely attributable to the passage of clouds over the sites that alternate with clear-sky periods. This type of sky condition is particularly hard for a medium-range forecast system, like \gls{NAM}, to simulate and predict. In fact, \gls{NAM} does not capture well the cloud dynamics on this particular day, producing a smooth time series peaking at 800 $W/m^2$, which corresponds to an irradiance value in the lower end of the distribution associated with clear-sky conditions. The optimal \gls{AnEn} is able to compensate for the high bias of \gls{NAM} predictions, leading to an increase in the prediction variability around the averaged observed value, although the observed variance still appears to be underestimated. \gls{DA}, on the other hand, is able to reconstruct the observed temporal variability and to better correct \gls{NAM}.

To better understand the method's performances, \Figs{fig:solar} (b, c) shows a subset of the original forecast variables and the latent features learned by \gls{DA} at 2 PM on August 8th, 2019, denoted by the shaded timestamp on \Fig{fig:solar} (a). The optimal \gls{AnEn} uses five predictors while \gls{DA} uses all 227 predictors available. Due to the limit of visualization, only solar irradiance and surface relative humidity are plotted in \Fig{fig:solar} (b) from the original forecast variable space; only the first two latent features are plotted in \Fig{fig:solar} (c) from the transformed latent space. The grey dots indicate all the historical forecasts in the search repository from 2011 to 2018 and they collectively show the distribution of these historical forecasts in the two-dimensional space. The target forecast, namely the \gls{NAM} forecast valid at 2 PM on August 8th, 2019, is shown as a red point. The historical forecasts selected by optimal \gls{AnEn} are denoted as orange crosses and the historical forecasts selected by \gls{DA} are denoted as blue pluses. The ideal members are the historical forecasts selected by the reverse analog process, denoted as green dots. These members are ideal in the sense that the corresponding observations to these forecasts will produce the most accurate predictions. In other words, these forecasts are associated with the observations that are closest in values to the true solar irradiance value (at around 460 $W/m^s$). An ideal analog similarity function would lead to selecting these 11 green dots as weather analogs, and then the resulted ensemble would center around the true value with a sharp distribution.

It is clear in the original forecast variable space, as in \Fig{fig:solar} (b), the green dots are scattered across different regions of the forecast distribution. Because a Euclidean distance is used to define similarity, the optimal \gls{AnEn} selected dots that are closest to the target forecast in the original space. Note that it might not seem as obvious from the figure because only two dimensions are plotted. In practice, there are five dimensions for the optimal \gls{AnEn}. This inability to choose good analogs could be caused by not having enough examples in the training for the standard AnEn algorithm, or that the metric or the selected set of predictors is not optimal. However, as shown in \Fig{fig:solar} (c), \gls{DA} first transformed the variable space into a 20-dimensional latent space, and then selected the closest candidates. Although none of the green dots (the ideal forecasts) are actually selected as analogs, they are much more clustered around the target forecast compared to the original space, making them more likely to be selected. This is likely the result of adopting an improved metric and set of predictors. Indeed, the difference in the similarity metric leads to the result that weather analogs selected by the optimal \gls{AnEn} are further apart in the latent space while the weather analogs selected by \gls{DA} are further apart in the original forecast variable space. This difference in the analog selection behaviors is causing the differences in the final predictions. It indicates that additional knowledge has been ingested into the transformation function employed by \gls{DA}. The improved performance benefits from the training stage of the triplet network with a reverse analog procedure. Similar behaviors have also been observed in wind speed predictions (not shown).

\begin{figure}
    \begin{minipage}{0.5\textwidth}
        \centering
        \includegraphics[width=\textwidth]{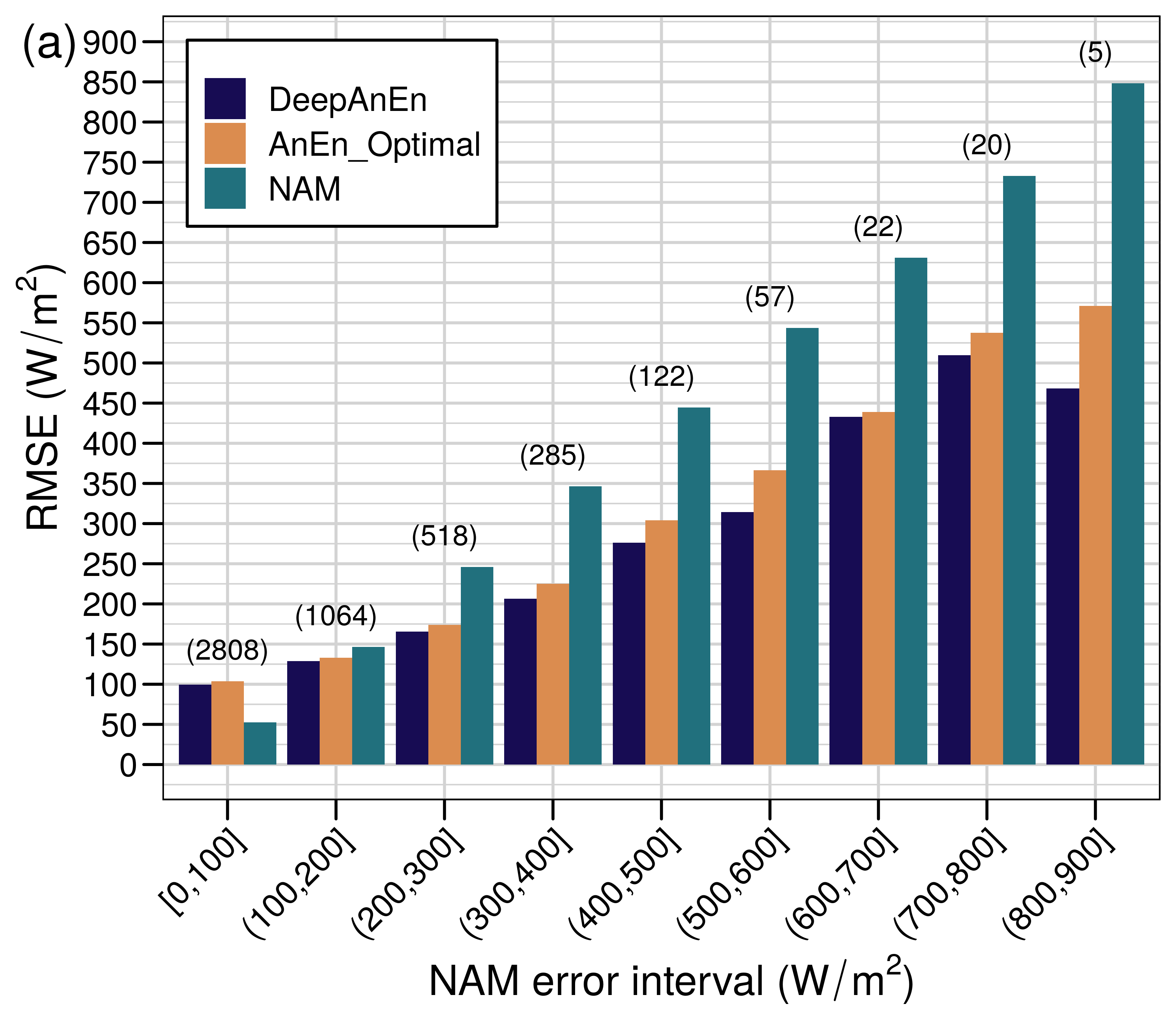}
        \phantomsubcaption
        \label{fig:solar_interval}
    \end{minipage}
    \begin{minipage}{0.5\textwidth}
        \centering
        \includegraphics[width=\textwidth]{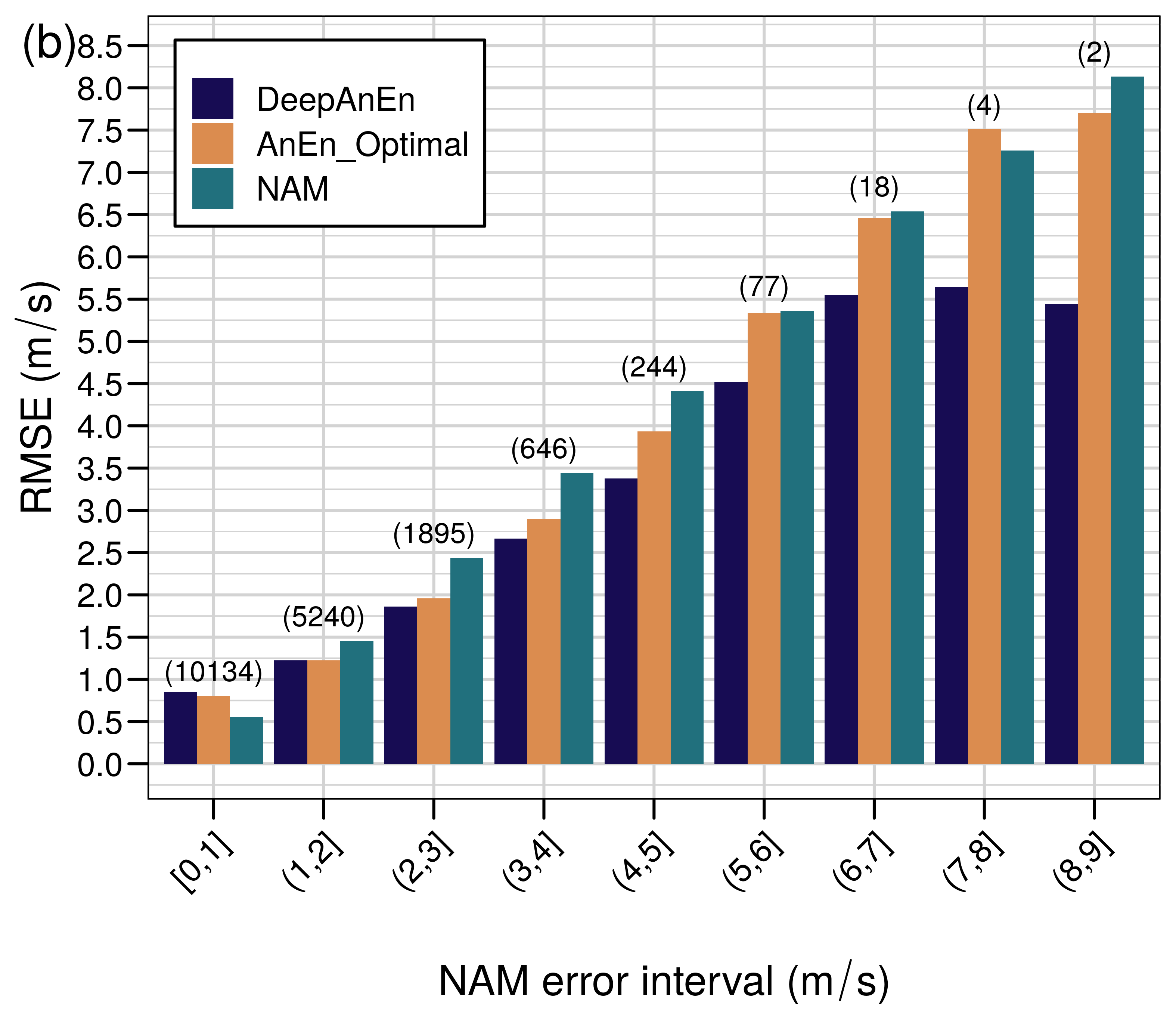}
        \phantomsubcaption
        \label{fig:wind_interval}
    \end{minipage}
    \caption{\gls{RMSE} for (a) solar irradiance and (b) wind speed forecasts binned with \gls{NAM} prediction error intervals. Verification for wind speed forecasts are done for all forecast lead times; verification for solar irradiance forecasts are only done for lead times valid during day times to prevent possible skewing of the results. The number on top of each bar group denotes the total number of forecasts verified in the particular error interval group.}
\end{figure}

To quantify the ability of \gls{DA} and the optimal \gls{AnEn} in reducing errors of different magnitudes, \gls{NAM} predictions are binned based on their individual forecast error. \Fig{fig:solar_interval} shows the \gls{RMSE} summary for solar irradiance forecasts. It is clear that analog-based ensemble techniques improve \gls{NAM} for all error intervals except for the lowest ones. At the same time, \gls{DA} consistently demonstrate greater capability of error correction compared to \gls{AnEn}, indicating by a smaller \gls{RMSE} compared to \gls{AnEn}. Similar results can be observed from \Fig{fig:wind_interval} on wind speed predictions. While the improvement from \gls{NAM} to \gls{AnEn} is not as great as in the case of solar irradiance predictions, \gls{DA} consistently offers better error correction to \gls{NAM} compared to \gls{AnEn}, leading to a lowest \gls{RMSE}, except for the lowest error interval group. Interestingly, neither analog-based ensemble technique beats the underlying \gls{NAM} model in the lowest error interval. This indicates that, if \gls{NAM} already provides an accurate deterministic prediction, the analog-based ensemble technique might decrease, rather than further increase, the prediction accuracy.


In summary, using more variables with an \gls{ML} driven similarity metric while identifying weather analogs has the advantage to correct predictions for cases where the underlying \gls{NWP} model produces significant errors, as shown by \gls{DA} greater ability to correct larger errors than the optimal \gls{AnEn}. This has a great implication for its application to hard-to-predict events where the \gls{NWP} model fails and the conventional \gls{AnEn} has been found to produce a systematic bias \cite{alessandrini_improving_2019}.

\subsection{Effect of Model Updates}
\label{sect:model-updates}

\begin{figure}[htp!]
    \centering
    \includegraphics[width=0.9\textwidth]{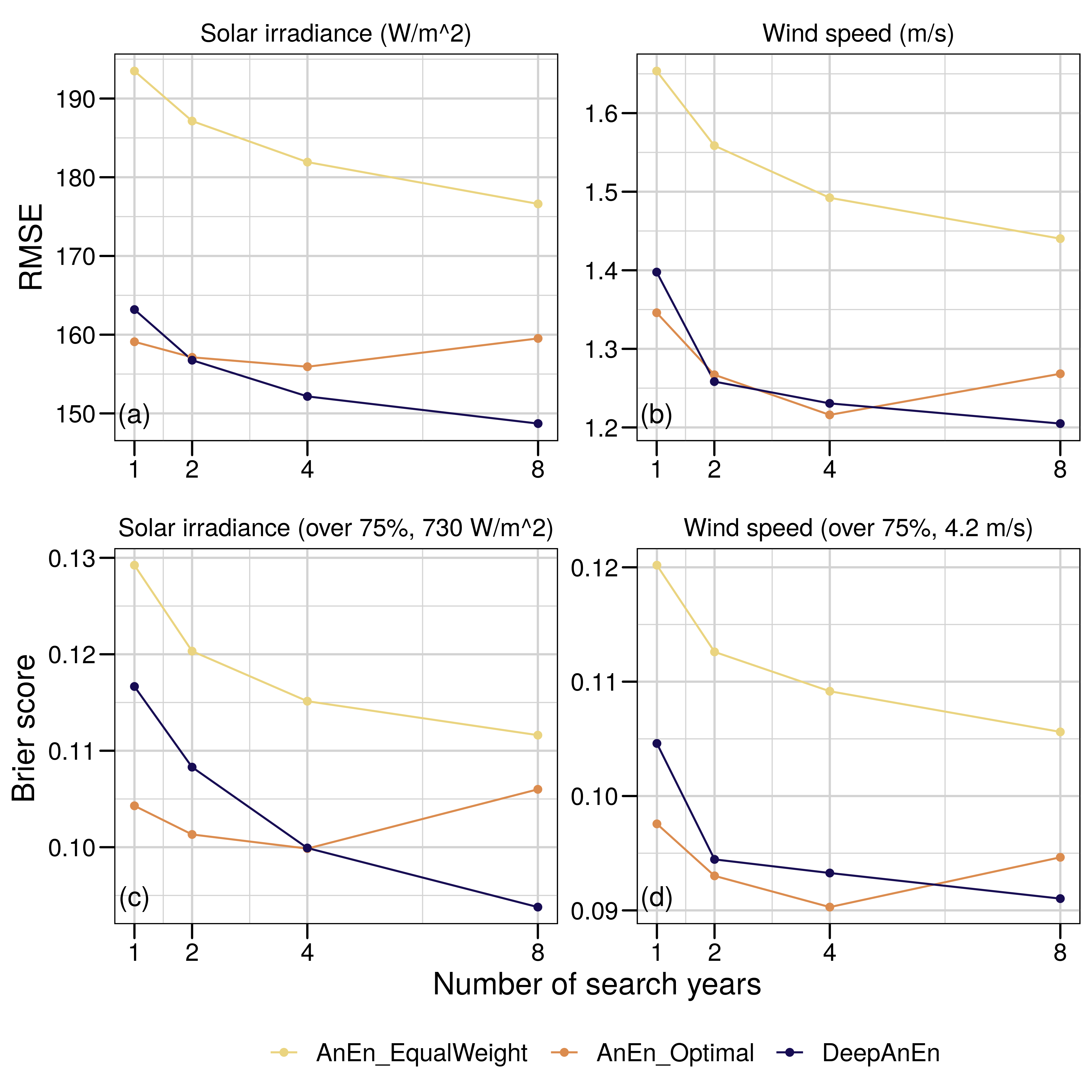}
    \caption{Sensitivity studies of the length of historical search repository for predicting solar irradiance and wind speed. Historical forecasts from 2018 are used for the single-year case; forecasts from 2017 to 2018 are used for the two-year case; forecasts from 2015 to 2018 are used for the four-year case; and forecasts from 2011 to 2018 are used for the eight-year case. (a, b) show the \gls{RMSE} and (c, d) show the Brier scores. The performance metrics are generated from a subset of lead times (11 AM to 12 PM) to prevent skewed verification due to night times. Weights used by the optimal \gls{AnEn} are optimized using the eight-year case; the embedding network used by \gls{DA} is trained with the eight-year case.}
    \label{fig:years}
\end{figure}

It is often assumed that better weather analogs can be found with a longer repository of historical forecasts, particularly for less frequent events. This assumption, however, can be violated whenever there are updates to the underlying \gls{NWP} model, which can potentially pose a limit on the performance of any postprocessing method built on that model. Depending on the nature of the updates, a long historical search repository of an operational model might not be as useful as one would initially expect. This section focuses on sensitivity of the length of the historical search repository and its impact on prediction errors.

\gls{NAM} has been constantly updated with major and minor changes. The most important change was an update of its core from the Eta model to the \gls{WRF} model in 2006. After that, \gls{NAM} has had roughly a minor update every several months \footnote{The update frequency is estimated from the changelogs from the official code base on GitHub at \url{https://github.com/wrf-model/WRF}}. There is currently little information on how tolerant \gls{AnEn} is in terms of model updates. Part of the reason lies within the intrinsic difficulty in quantifying the tolerance because the impact of a model update on the model itself can be highly complex and sometimes even unknown. Quantifying the impact on \gls{AnEn} performance would add another layer of complexity.

This paper shows results using two common metrics, \gls{RMSE} and the Brier score, comparing the performance of analog-based methods with different lengths of search repository. Results do not focus on causal relationship but only examine the sensitivity of various analog-based methods.

\Figs{fig:years} show the verification for solar irradiance and wind speed predictions with different numbers of historical years in the search repository. The Brier score quantifies how well the analog-based methods predict an event over the 75\%-percentile value of the observed distribution. This 75\%-percentile value corresponds to 730 $W/m^2$ in irradiance and 4.2 $m/s$ in wind speed. To prevent possible skew of verification since solar irradiance is only abundant during the daytime, results are calculated the local time from 11 AM to 12 PM over the entire test period of 2019.

All four panels in \Fig{fig:years} demonstrate a similar pattern when increasing the number of historical years in the search repository. The optimal \gls{AnEn} (orange solid line) has a ``saturation point'' at around four years where it then stops decreasing its forecast errors given more historical forecasts. Its error starts to increase given more than four years of historical forecasts. Interestingly, the equally weighted \gls{AnEn} with 227 input forecast variables, shown in solid yellow lines, continues to decrease when more historical data are added, although the error remains high. \gls{DA}, shown in solid blue lines, has a similar pattern that continues to decrease forecast errors given more historical forecasts while achieving significantly better results than the equally weighted \gls{AnEn}. The consistent trend of the error decreasing for \gls{DA} suggests that it is able to account for more degrees of freedom in the forecasts and that additional data allows to find more robust latent features throughout multiple years even when the \gls{NWP} model has gone through routine updates. However, it is not definitive that the ``saturation point'' for regular \gls{AnEn} is at four years because the magnitude of error changes is relatively small. It has only been 11 years since \gls{NAM} changed its model core in 2009, so a longer data repository from a different model is needed to study this problem in a more conclusive fashion.

\subsection{Prediction Spatial Patterns}

Previous sections have evaluated the three analog-based methods (\gls{DA}, the optimal \gls{AnEn}, and the equally weighted \gls{AnEn}) at Pennsylvania State University. To investigate the effectiveness of \gls{DA} on a spatial domain, this section extends the geographic scale from Pennsylvania State University to the entire region of Pennsylvania. The area of Pennsylvania, consisting of 1225 grid locations, is selected for its direct vicinity to our previously studied area. Analogs are sought independently at each location and the \gls{ML} models are trained with the same architecture. Prediction errors are calculated with regard to \gls{NAM} model analysis initiated at later forecast cycles, 1800 UTC, due to the lack of a continuous spatial coverage from \gls{SURFRAD} networks. \gls{RMSE} is average across the test period of August 2019, and four forecast lead times valid at 2, 3, 4, 5 PM local time. This is primarily to test the performance during daytime when larger variations in both solar irradiance and wind speed are to be expected.

\begin{table}[t]
    \centering
    \scalebox{0.8}{
    \begin{tabular}{ccccc}
        \hline
        \textbf{Experiment} & \textbf{Predictand} & \textbf{Vertical Level} & \textbf{Trained ML Models} & \textbf{Training Data Sites} \\ \hline
        1                   & Solar irradiance    & Surface                 & 1                          & 1                            \\
        2                   & Solar irradiance    & Surface                 & 1                          & 100                          \\
        3                   & Wind speed          & Surface                 & 1                          & 1                            \\
        4                   & Wind speed          & Surface                 & 1                          & 100                          \\
        5                   & Wind speed          & Surface                 & 8                          & 408                          \\
        6                   & Wind speed          & 80 meters above ground  & 1                          & 1                            \\
        7                   & Wind speed          & 80 meters above ground  & 1                          & 100                          \\
        8                   & Wind speed          & 80 meters above ground  & 9                          & 408                          \\
        \hline
    \end{tabular}}
    \caption{A summary of experiments for training configurations of \gls{ML} models for spatial predictions of \gls{DA}}
    \label{tab:space-experiments}
\end{table}

\begin{table}[t]
\centering
\scalebox{0.8}{
\begin{tabular}{cccc}
\hline
\textbf{Variable} & \textbf{Configuration} & \textbf{Average RMSE (Min / Max)} & \textbf{Improvement (\%)} \\ \hline
\multirow{5}{*}{Solar irradiance ($W/m^2$)} & NAM & 158.02 (136.59 / 182.06) & * \\
 & AnEn\_EqualWeight & 162.43 (145.33 / 182.16) & -2.79 \\
 & AnEn\_Optimal & 133.66 (119.47 / 151.88) & 15.42 \\
 & DeepAnEn 1 @ 1 & 136.63 (120.87 / 161.42) & 13.54 \\
 & DeepAnEn 1 @ 100 & {\ul \textbf{130.31 (111.24 / 148.01)}} & {\ul \textbf{17.54}} \\ \hline
\multirow{6}{*}{Surface wind speed ($m/s$)} & NAM & 0.81 (0.61 / 1.12) & * \\
 & AnEn\_EqualWeight & 1.06 (0.78 / 1.68) & -30.76 \\
 & AnEn\_Optimal & 0.75 (0.54 / 1.07) & 7.56 \\
 & DeepAnEn 1 @ 1 & 0.83 (0.64 / 1.62) & -2.62 \\
 & DeepAnEn 1 @ 100 & {\ul \textbf{0.68 (0.51 / 1.07)}} & {\ul \textbf{15.74}} \\
 & DeepAnEn 8 @ 408 & 0.7 (0.52 / 1.09) & 13.27 \\ \hline
\multirow{6}{*}{80-meter wind speed ($m/s$)} & NAM & 1.09 (0.91 / 1.46) & * \\
 & AnEn\_EqualWeight & 1.46 (1.16 / 2) & -34.08 \\
 & AnEn\_Optimal & 1.01 (0.81 / 1.32) & 7.47 \\
 & DeepAnEn 1 @ 1 & 1.09 (0.87 / 1.8) & 0.00 \\
 & DeepAnEn 1 @ 100 & {\ul \textbf{0.94 (0.77 / 1.33)}} & {\ul \textbf{13.21}} \\
 & DeepAnEn 9 @ 408 & 0.95 (0.75 / 1.33) & 13.07 \\ \hline
\end{tabular}}
\caption{A summary of results from various methods. \textit{DeepAnEn 1 @ 1} represents \gls{DA} predictions with 1 \gls{ML} model trained with data from 1 site.}
\label{tab:space-results}
\end{table}

\Table{tab:space-experiments} summarizes the different configurations of \gls{ML} models for spatial predictions of \gls{DA}. A series of experiments have been carried out specifically to investigate

\begin{enumerate}
    \item the performance of \gls{DA} on three weather variables,
    \item the impact of including more training data from nearby locations,
    \item and the potential benefit of training more than one \gls{ML} model for \gls{DA}.
\end{enumerate}

When only one training data site is used (\textit{Exp. 1, 3, and 6}), the model grid closest to the \gls{SURFRAD} station at Pennsylvania State University is used. The \gls{ML} model is trained using data from only this grid point and then applied to the entire Pennsylvania. When 100 training data sites are used (\textit{Exp. 2, 4, and 7}), these sites are equally spaced across Pennsylvania. \textit{Exp. 5 and 8} have used a total of 408 training data sites and trained several \gls{ML} models. The sites and the number of models are determined as follows: the annual wind speed for each grid point is first calculated and then, all grid points are classified and binned with a 0.5 $m/s$ interval; for each interval, an \gls{ML} is trained using data from the model grids belonging to this interval. Due to the machine memory limit, the maximum number of training data sites for a particular \gls{ML} is 100. When the total number of grid pints belonging to a particular interval exceeds this limit, 100 sites are randomly selected from the grid points.

\Table{tab:space-results} shows the average \gls{RMSE} from various forecast methods across the entire domain. The test period is 2019 daily from 2 to 5 PM. The night time is excluded from the verification period because of the reduced variation and amount of solar irradiance and wind speed during nights. Verification of the entire daytime period is unfortunately impossible because, due to data management and archive limit, \gls{NAM} does not provide the analysis field for a continuous daytime verification. Wind speed verification is carried out for strong wind cases ($> 4 m/s$). For all three variables, the equally weighted \gls{AnEn} generally has a higher error than the baseline \gls{NAM} predictions. This is because weight optimization is essential to the performance of the \gls{AnEn} technique. In the cases of the optimal \gls{AnEn}, when the number of predictor variables has been reduced and the respective weights have been optimized using historical data, the average \gls{RMSE} shows consistent drops compared with the baseline \gls{NAM} predictions, 15.42\% for solar irradiance, 7.56\% for surface wind speed, and 7.47\% for 80-meter wind speed. When comparing \gls{DA} and the optimal \gls{AnEn}, the single \gls{ML} model trained on 100 equally spaced sites consistently yields the best results, having the lowest \gls{RMSE} out of all alternative forecasts. When comparing multiple configurations of \gls{DA}, the \gls{ML} models trained on 100 locations are generally superior to the models trained on a single location. However, the expected improvement from using a mixture of \gls{ML} models (\textit{DeepAnEn 8 @ 408} and \textit{DeepAnEn 9 @ 408}) is not observed for wind speed predictions. In these experiments, when more than one \gls{ML} model is used by \gls{DA}, each model is responsible of predicting for a subset of the domain that has similar wind speed climatology. This chunking mechanism might be deemed inefficient for choosing the appropriate models. It implies that consuming more computation to train models might not necessarily help to further improve the performance of \gls{DA}. A different chunking mechanism that divides the domain considering both the physical variable (e.g., wind speed) and orography should be further investigated.

\begin{figure}[htp!]
    \centering
    \begin{minipage}{0.9\textwidth}
        \includegraphics[width=\textwidth]{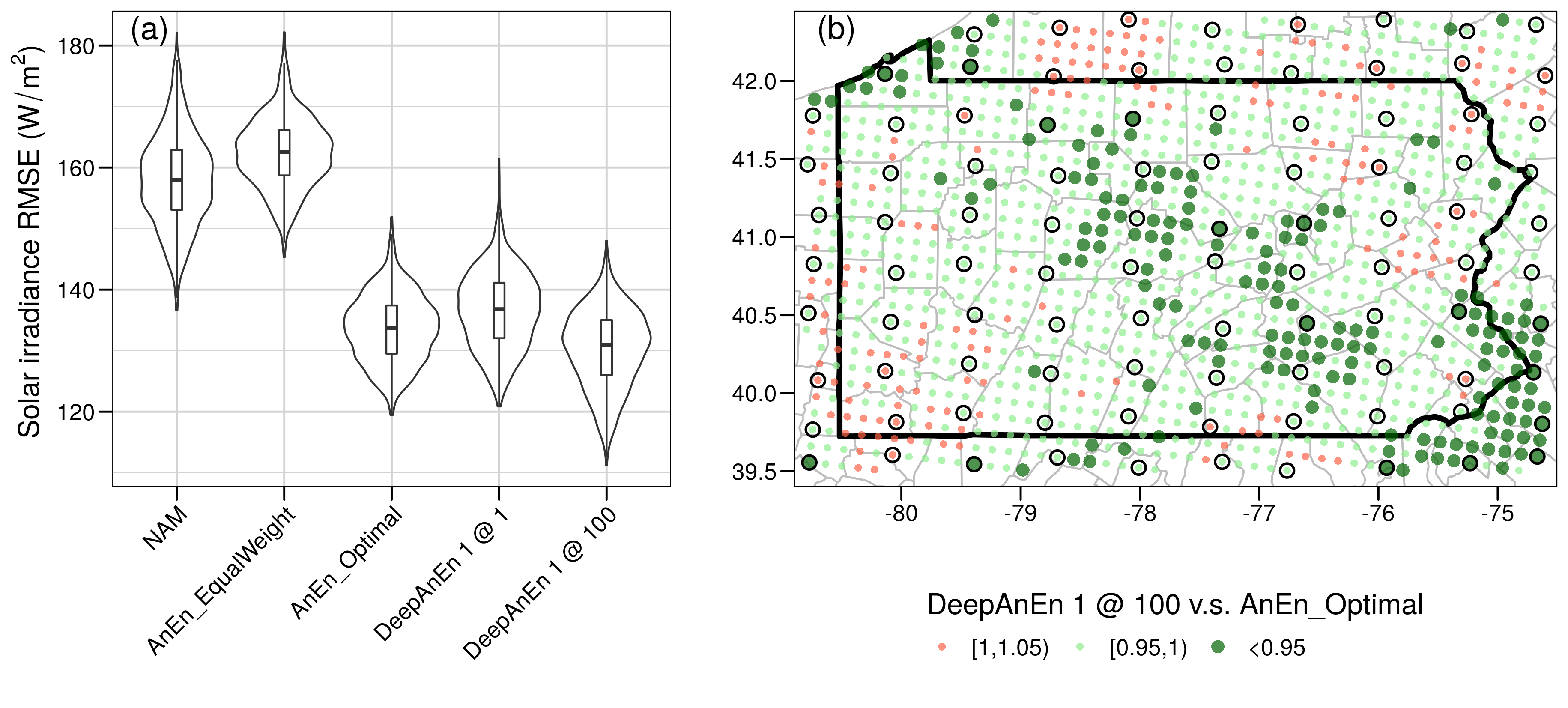}
    \end{minipage}
    \begin{minipage}{0.9\textwidth}
        \includegraphics[width=\textwidth]{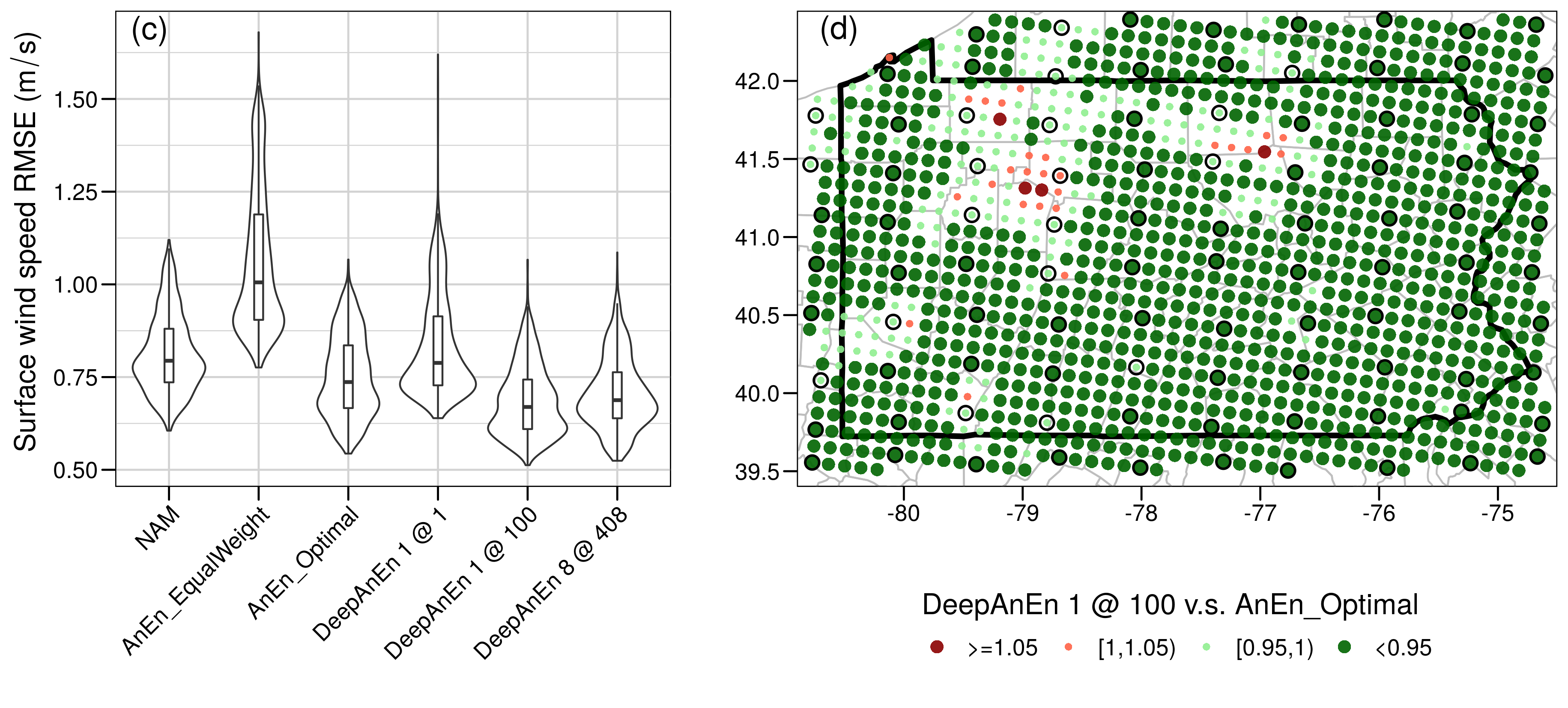}
    \end{minipage}
    \begin{minipage}{0.9\textwidth}
        \includegraphics[width=\textwidth]{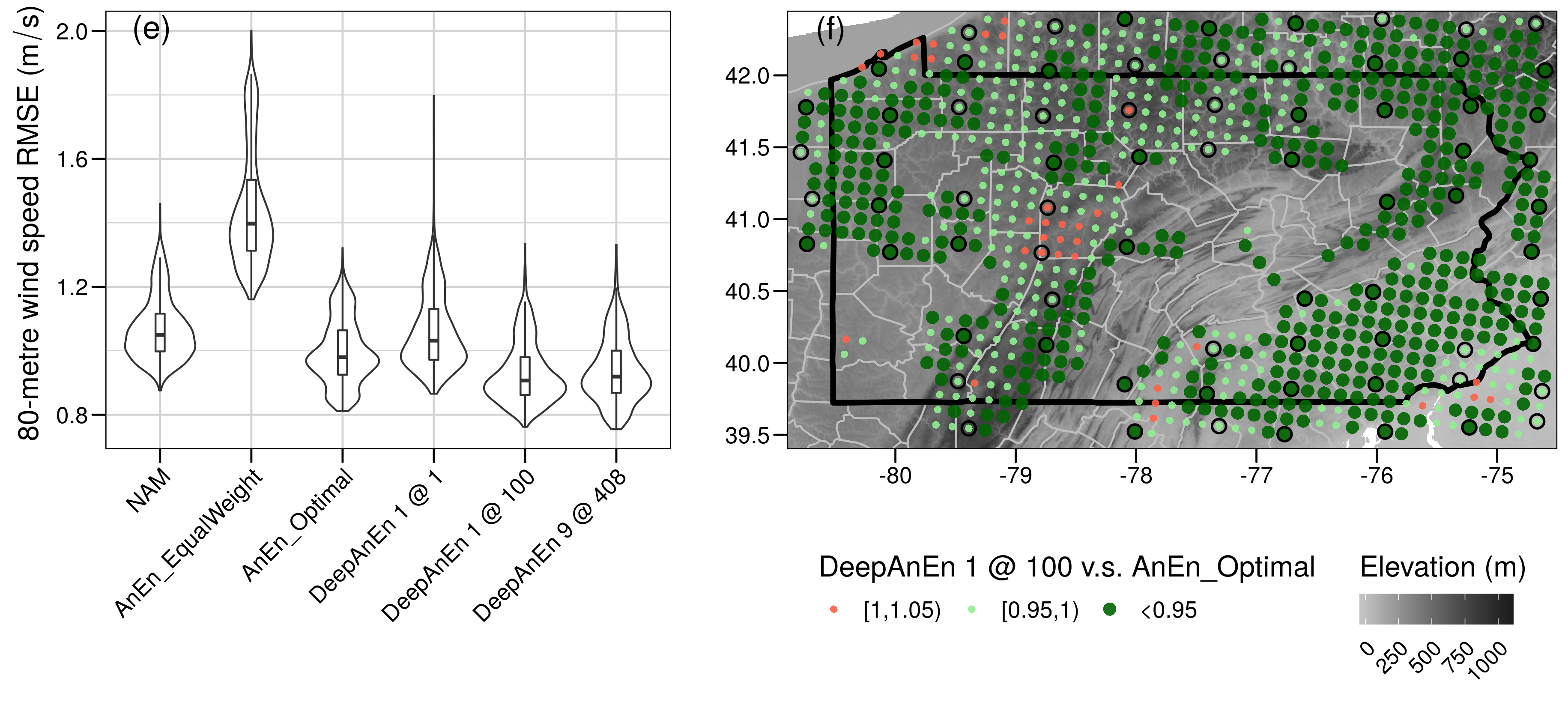}
    \end{minipage}
    \caption{Distributions of prediction errors (a, c, e) and geographic maps comparing \gls{DA} and the optimal \gls{AnEn} (b, d, f). The verification period is 2019 daily from 2 to 5 PM to favor the daytime period and only large wind (> 4 $m/s$) cases are verified. Predictions over water (upper-left in b, d, f) have been removed to focus predictions on land. In (a, c, e), the three hinges are, in turn, the first, second, and third quartiles; the whiskers extend from the hinge to the value at 1.5 * inter-quantile range of the hinge. Locations with an annual wind speed over 4 $m/s$ are shown in (f) to highlight strong wind region. Training data sites are circled in black.}
    \label{fig:space}
\end{figure}

\Figs{fig:space} (a, c, e) show the distributions of prediction errors from different methods and configurations. \Fig{fig:space} (a) shows the error distribution of solar irradiance predictions. \gls{NAM} and the equally weighted \gls{AnEn} have similar errors while the optimal \gls{AnEn} and \gls{DA} show reduced errors. Out of the five methods, \gls{DA} with 100 training data sites yield the best results. The long tail of large errors produced by \textit{DeepAnEn 1 @ 1} suggests that training an \gls{ML} model using only one data site could lead to a model ignorance on the spatial variation in the domain. This problem can be addressed by increasing the training data and the number of training data sites, as shown by \textit{DeepAnEn 1 @ 100}. Similar patterns can be observed from \Figs{fig:space} (b, c) on wind speed forecasts. \Fig{fig:space} (b) shows the the verification on surface wind speed over 4 $m/s$ and \Fig{fig:space} (c) shows the the verification on 80-meter wind speed over 4 $m/s$. Again, the equally weighted \gls{AnEn} failed to improve the baseline \gls{NAM} model while the optimal \gls{AnEn} reduces the prediction error. \gls{DA} with the model trained on 100 data sites doubles the improvements of the optimal \gls{AnEn}, producing 15.74\% for surface wind speed forecasts and 13.21\% for 80-meter wind speed forecasts.  

\Figs{fig:space} (b, d, f) show the geographic maps of \gls{RMSE} comparing \textit{DeepAnEn 1 @ 100} with the optimal \gls{AnEn}. Green indicates \gls{DA} has a lower error than \gls{AnEn} while red indicates \gls{DA} has a higher error than \gls{AnEn}. Dots are slightly enlarged when this difference is bigger than 5\% of the error of \gls{AnEn}. The depicted \gls{DA} uses an \gls{ML} model trained on 100 locations, circled in black. \Fig{fig:space} (f) only shows the verification at the locations where the annual wind speed at 80 meters above ground exceeds 4 $m/s$. These regions are favorable because of the wind abundance and their potential for wind farm investment. The predominant green region shows a consistent outperformance of \gls{DA} over the optimal \gls{AnEn} across a spatial domain. \gls{DA} reduces the prediction error for solar irradiance and wind speed when both resources are abundant temporally and spatially. This indicates \gls{DA} is suitable as an improved version of \gls{AnEn} implementation for renewable energy resource forecasts.

\begin{sidewaysfigure}[htp!]
    \centering
    \begin{minipage}{0.9\textwidth}
        \includegraphics[width=\textwidth]{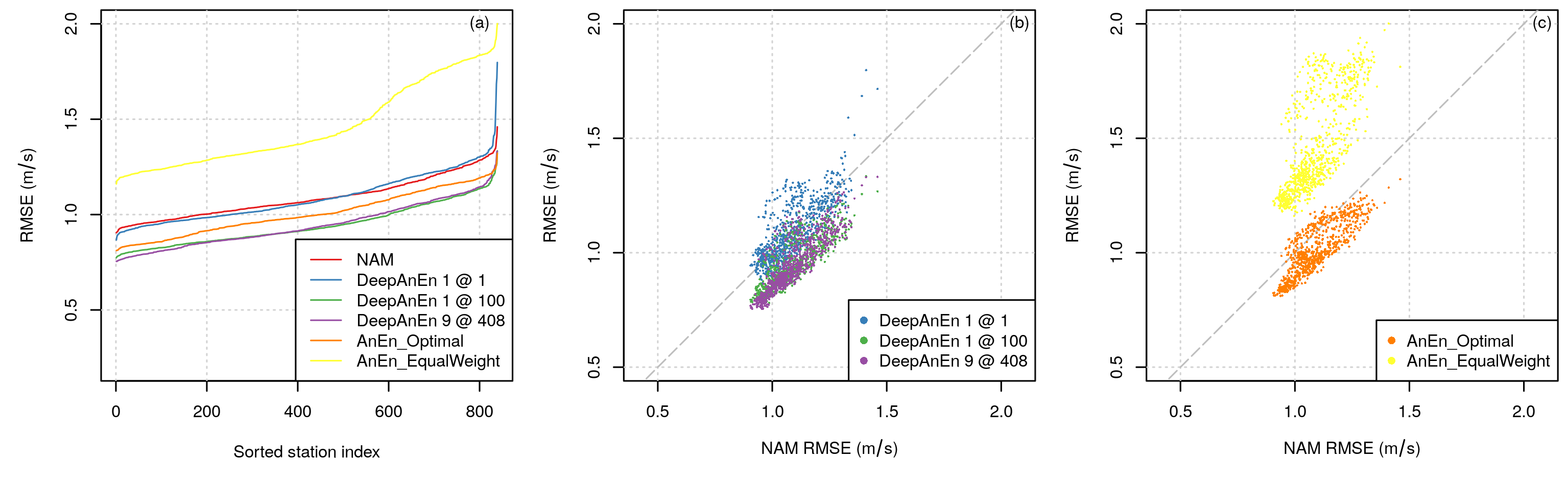}
    \end{minipage}
    \begin{minipage}{0.49\textwidth}
        \includegraphics[width=\textwidth]{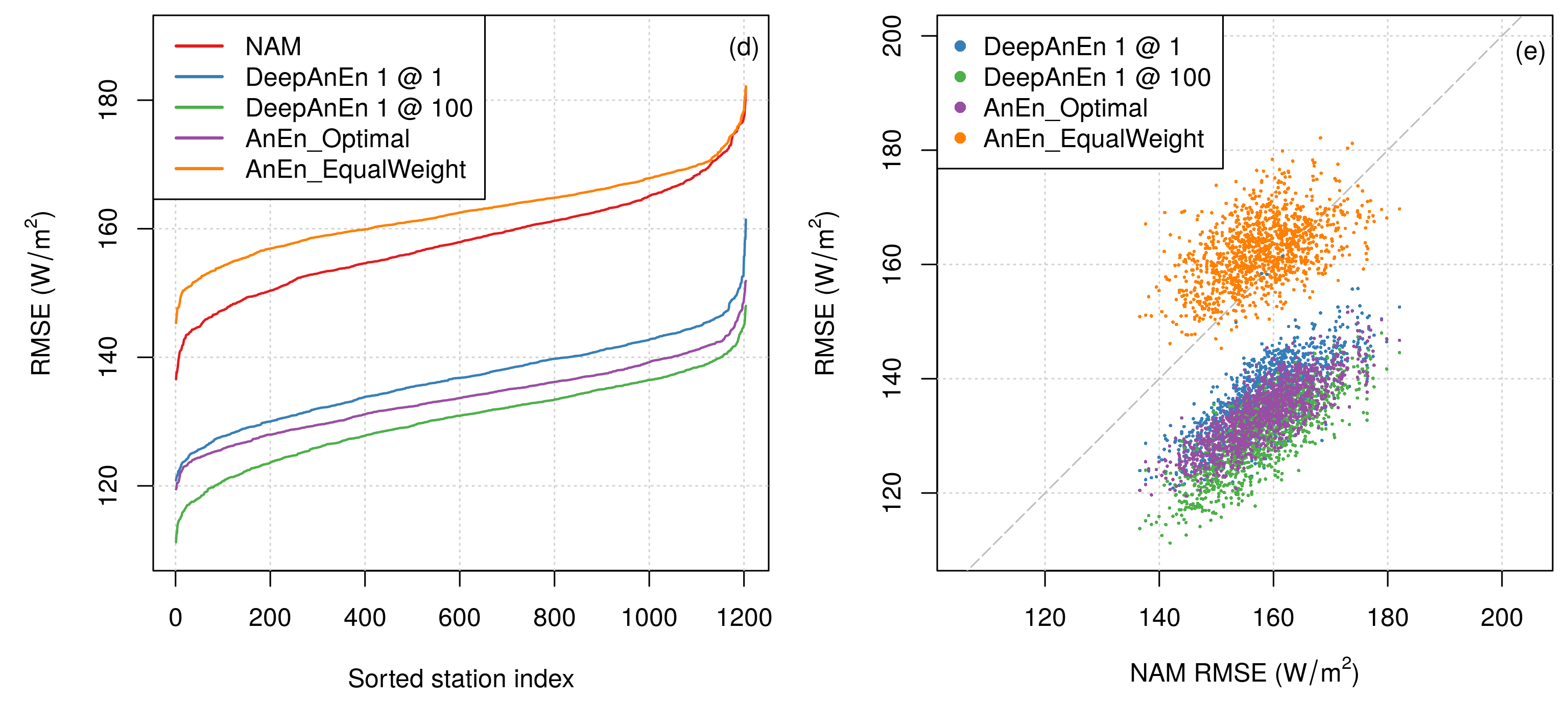}
    \end{minipage}
    \begin{minipage}{0.49\textwidth}
        \includegraphics[width=\textwidth]{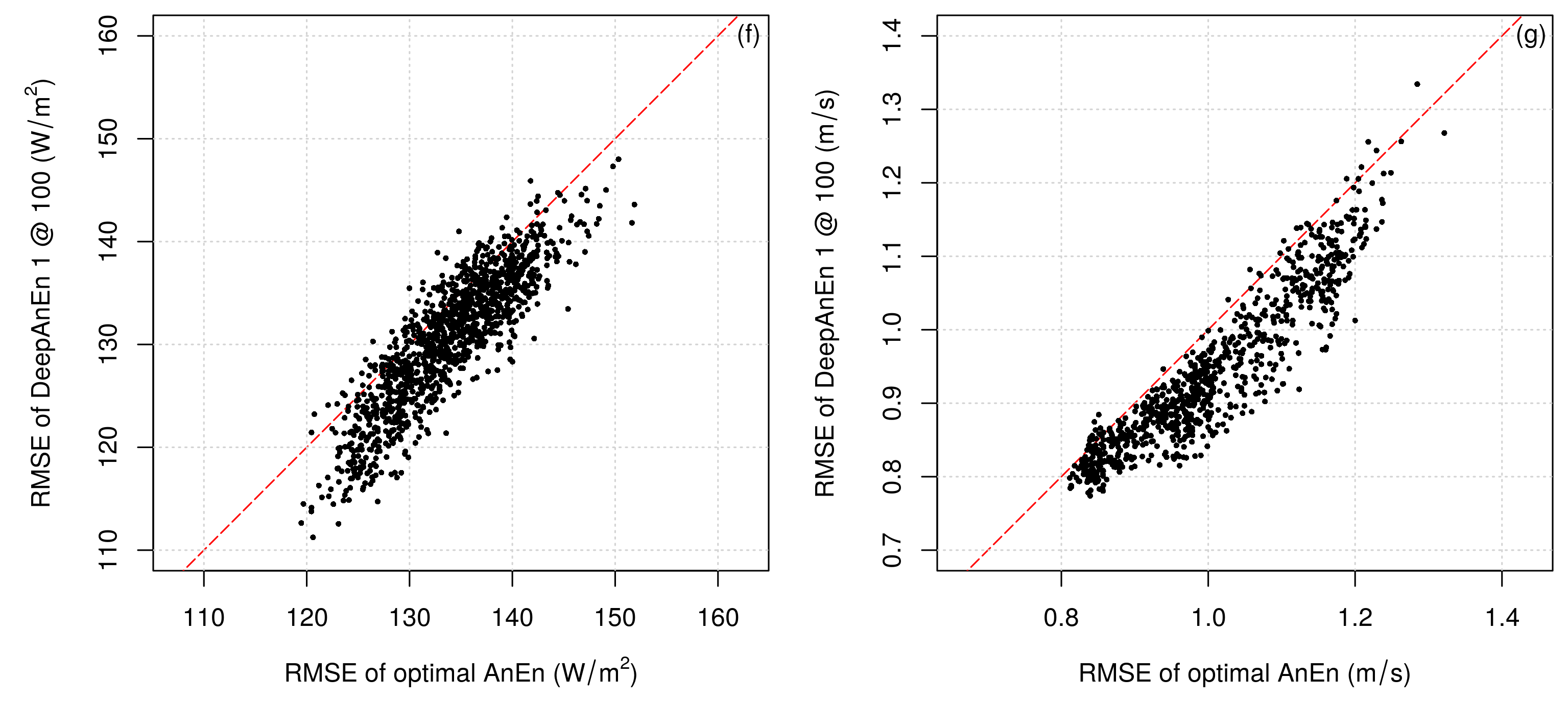}
    \end{minipage}
    \caption{\gls{RMSE} comparison for spatial predictions between various methods and configurations. (a, d) show the distribution of \gls{RMSE} and compare the sorted \gls{RMSE} from each location. (b, c) show the \gls{RMSE} scatter plots for wind speed forecasts at 80 meters above ground; (e) shows the \gls{RMSE} scatter plot for solar irradiance forecasts. (f, g) compares the \gls{RMSE} from the optimal \gls{AnEn} and \textit{DeepAnEn 1 @ 100}, for 80-meter wind speed (f) and solar irradiance (g).}
    \label{fig:space-points}
\end{sidewaysfigure}

To further illustrate the difference among the equally weighted \gls{AnEn}, the optimal \gls{AnEn}, and \gls{DA}, \gls{RMSE} for solar irradiance and 80-meter wind speed forecasts are compared in \Fig{fig:space-points}. \Fig{fig:space-points} (a) shows the sorted \gls{RMSE} at each location for 80-meter wind speed. \gls{RMSE} of the equally weighted \gls{AnEn} is much higher than the \gls{RMSE} of \gls{NAM} and \textit{DeepAnEn 1 @ 1}. In this case, the \gls{ML} model trained on only one location does not improve \gls{DA} predictions compared to \gls{NAM}. However, the optimization of using a trained model shows a great preference compared to the equally weighted \gls{AnEn}. The \gls{ML} model, although only trained on one location, is still able to learn the relative importance of 383 forecast variables from \gls{NAM} and to significantly reduce prediction errors compared to no optimization on weights at all. Going further down is the optimal \gls{AnEn} and then other two configurations of \gls{DA}, \textit{DeepAnEn 1 @ 100} and \textit{DeepAnEn 9 @ 408}. The two configurations of \gls{DA} have similar \gls{RMSE}, both lower than the optimal \gls{AnEn}, suggesting \gls{DA} with the model trained on multiple locations has the ability to outperform the optimal \gls{AnEn}. Note that training nine individual models would require nine times as much computation as for training one model, given similar amount of training data. However, this additional computation does not necessarily lead to the expected gain in prediction accuracy. \Figs{fig:space-points} (b, c) shows the \gls{RMSE} scatter plot using the verification results from \Fig{fig:space-points} (a). \textit{DeepAnEn 1 @ 1} (in blue) mostly lie on top of the diagonal line, indicating its similarity in \gls{RMSE} with \gls{NAM}. The equally weighted \gls{AnEn} (in yellow) lies high above the diagonal line suggesting the low prediction accuracy. Point clouds of \textit{DeepAnEn 1 @ 100}, \textit{DeepAnEn 9 @ 408}, and the optimal \gls{AnEn} are clustered below the diagonal line, suggesting better accuracy than the baseline \gls{NAM} predictions. 

\Figs{fig:space-points} (d, e) show the \gls{RMSE} on solar irradiance forecasts following the same argument. \Fig{fig:space-points} (d) shows the sorted \gls{RMSE} at each location and \Fig{fig:space-points} (e) compares the \gls{RMSE} scatter plots. There are generally two clusters, one from \gls{NAM} and the equally weighted \gls{AnEn} and another one from \textit{DeepAnEn 1 @ 1}, \textit{DeepAnEn 1 @ 100}, and the optimal \gls{AnEn}. In the second cluster, there is a clear cascade in the sorted \gls{RMSE} that \textit{DeepAnEn 1 @ 100} outperforms the optimal \gls{AnEn} and then the \textit{DeepAnEn 1 @ 1}. This is supported by the vertical shift of cloud points on \Fig{fig:space-points} (e), from green, purple, to blue. These experiments demonstrate the effectiveness of \gls{DA} in predicting solar irradiance.

Finally, \Figs{fig:space-points} (f, g) specifically compare \gls{RMSE} between the best configuration of \gls{DA}, \textit{DeepAnEn 1 @ 100}, and the optimal \gls{AnEn}. In both figures, most of points lie below the diagonal line suggesting \gls{DA} has a lower \gls{RMSE} and therefore a higher accuracy in predicting solar irradiance and 80-meter wind speed. Specifically, 85.88\% (1034 out of 1204) of the points in \Fig{fig:space-points} (f) and 95.95\% (805 out of 839) in \Fig{fig:space-points} (g) lie below the referential line. These results show the ability of \gls{DA} to predict day time solar irradiance and wind speed variation for a spatial domain, achieving an improved accuracy compared to the optimal \gls{AnEn}.

\section{Discussion and Conclusions}
\label{sect:conclude}

This paper introduces a new technique of generating weather analogs using \gls{ML}.  Specifically, weather analogs are sought in a transformed space generated by a pre-trained \gls{ML} model, and as a result, the similarity metric is renovated and redefined in the transformed space. We, hereby, highlight some of the major motivation and findings for the \gls{ML} driven \gls{AnEn} technique:

\begin{enumerate}
    \item The conventional similarity metric \cite{delle_monache_probabilistic_2013} calculates weighted Euclidean distances in the original predictor variable space. Due to this practice, \gls{AnEn} is subject to using a few predictor variables and a computationally expensive weight optimization process. \gls{DA} overcomes the limits on the number of predictors by training an \gls{ML} model for feature transformation and representation.
    \item Optimizing predictor weights for the conventional \gls{AnEn} is found to be even more difficult for generating spatial predictions or gridded forecast output \cite{junk_predictor-weighting_2015, clemente2016analog}. However, \gls{DA} is able to generate more accurate predictions than the optimal \gls{AnEn} over a spatial domain. This is largely due to the powerful \gls{ML} model architecture trained using data from multiple sites.
    \item Using more weather variables to define weather analogs offers greater flexibility to various weather regimes, and it is more robust to model forecast errors. \gls{DA} has been found to be more accurate in predicting hard-to-forecast cases compared to \gls{AnEn} using only a few predictor variables.
    \item Finally, \gls{DA} has also been found to be more tolerant to model updates when increasing the historical archive for weather analog search. During model training using the proposed reverse analog technique, \gls{DA} builds a relationship between model forecasts and the analysis error, and later on, relies on this relationship to generate informed predictions. 
\end{enumerate}


The triplet training architecture is another key to learn effective weather features. \citeauthor{fanfarillo2020probabilistic} proposed to use a encoder-decoder architecture for analog generation to save computation and memory. They achieve constant scaling in computation and memory when the size of model archive increases. But yet the generative model was not able to achieve as good results as \gls{AnEn}. This work, however, is able to show that \gls{DA} outperforms the optimal \gls{AnEn} with the original similarity metric. This suggests that the features learned by the \gls{ML} embedding network are more powerful than identifying analogs in the original predictor variable space or in the latent space learned by an auto-encoder. There are two potential reasons: (1) auto-encoder, acting like a data compressor, seeks to represent the complex \gls{NWP} model forecasts with a few features. The \gls{NWP} model, however, is highly complex. And there is no guidance when constructing the latent features that will help to generate accurate ensemble forecasts. The only guidance is to represent the \gls{NWP} model forecasts. The quality of the latent features is yet to be ensured; (2) the features generated by an auto-encoder might not be continuous in the latent space. Clustered points in the latent space might not necessarily correspond to actually similar weather forecasts. Triplet networks, however, were specifically designed to overcome these issues. Its combination with a contrastive loss function particularly encourages to cluster points if they correspond to actual similar features in the original space. These clustered points can then contribute to an ensemble with better quality and improved accuracy (as shown in \Fig{fig:solar}). The reverse analog technique, introduced in \Sect{sect:learning}, offers specific guidance during the triplet network training, to learn features that would consider the final ensemble prediction and compensate for model errors.

\citeauthor{van1994searching} once estimated that $10^{30}$ years are needed to find good analogs. This estimation, fortunately, has been significantly reduced to several years using new analog theories. In general, it was assumed that better analogs can be found given a longer search history and these better analogs can lead to better predictions and improved ensembles. This work, however, shows that with the current implementation of \gls{AnEn}, the similarity metric is limited in flexibility and model update tolerance. \gls{DA} offers a good solution to these problems, yet a timely one because of the recent advancement in earth observation, weather modeling, and high-performance computing. Long archives of historical model simulations and observations are becoming easily accessible and \gls{DA} shows a great capability to harvest the recent advancement in weather analog identification. Future research should be directed to the application of such a framework in extreme weather forecasting and forecasts over a large spatial domain. \gls{LSTM} is used as the embedding network in this work, but studies are also encouraged to systematically evaluate the performance of different \gls{NN}s as the embedding network.

\bibliographystyle{unsrtnat}
\bibliography{bib.bib}

\end{document}